\documentclass[fleqn,usenatbib]{mnras}

\usepackage{newtxtext,newtxmath}

\usepackage[T1]{fontenc}

\DeclareRobustCommand{\VAN}[3]{#2}
\let\VANthebibliography\thebibliography
\def\thebibliography{\DeclareRobustCommand{\VAN}[3]{##3}\VANthebibliography}

\usepackage{graphicx}	
\usepackage{amsmath}	
\usepackage{subcaption}
\usepackage{multicol}

\title[]{A field-level emulator for modeling baryonic effects across hydrodynamic simulations}

\author[]{
Divij Sharma,$^{1, 2}$\thanks{E-mail: divijsharma@berkeley.edu}
Biwei Dai,$^{1, 2}$
Francisco Villaescusa-Navarro,$^{3, 4}$
and 
Uroš Seljak$^{1, 2}$
\\
$^{1}$ Berkeley Center for Cosmological Physics and Department of Physics, University of California, Berkeley, CA 94720, USA\\
$^{2}$ Lawrence Berkeley National Lab, 1 Cyclotron Road, Berkeley, CA 94720, USA\\
$^{3}$ Center for Computational Astrophysics, Flatiron Institute, 162 5th Avenue, 10010, New York, NY, USA\\
$^{4}$ Department of Astrophysical Sciences, Princeton University, 4 Ivy Lane, Princeton, NJ 08544 USA
}

\date{Accepted XXX. Received YYY; in original form ZZZ}

\pubyear{2023}

\begin{document}
\label{firstpage}
\pagerange{\pageref{firstpage}--\pageref{lastpage}}
\maketitle

\begin{abstract}
We develop a new and simple method to model baryonic effects at the field level relevant for weak lensing analyses. We analyze thousands of state-of-the-art hydrodynamic simulations from the CAMELS project, each with different cosmology and strength of feedback, and we find that the cross-correlation coefficient between full hydrodynamic and N-body simulations is very close to 1 down to $k\sim10~h{\rm Mpc}^{-1}$. This suggests that modeling baryonic effects at the field level down to these scales only requires N-body simulations plus a correction to the mode's amplitude given by: $\sqrt{P_{\rm hydro}(k)/P_{\rm nbody}(k)}$. In this paper, we build an emulator for this quantity, using Gaussian processes, that is flexible enough to reproduce results from thousands of hydrodynamic simulations that have different cosmologies, astrophysics, subgrid physics, volumes, resolutions, and at different redshifts. Our emulator is accurate at the percent level and exhibits a range of validation superior to previous studies. This method and our emulator enable field-level simulation-based inference analyses and accounting for baryonic effects in weak lensing analyses.

\end{abstract}




\section{Introduction}\label{sec:Introduction}
Weak gravitational lensing is a powerful tool for measuring 
the clustering of matter in our universe, and thus obtaining
information about matter content and initial conditions of our universe \citep{Kilbinger_2015} through various summary statistics. However, deriving precise cosmological constraints from weak lensing observations necessitates highly accurate theoretical models that account for baryonic physics, which redistributes matter on small scales via processes such as Active Galactic Nuclei (AGN) feedback. These processes remain poorly understood and inadequately constrained by current observations, leading to challenges in formulating a predictive theory.

While causality ensures that baryonic effects are negligible on large scales \citep{Lewis_2011}, baryonic effects gain significance on smaller scales, affecting structure formation through hydrodynamic processes. These processes, including AGN and stellar feedback, can heat gas and inject large amounts of energy into galaxies and the surrounding halo.  Specifically, AGN feedback can eject gas to very large distances, which can further modify the dark matter distribution through gravitational interactions.  

Studies have shown that probing the small scales contains a wealth of information, leading to stronger parameter constraints \citep{Lu_2021}, while ignoring small scale information leads to significant deterioration of these constraints \citep{Kohlinger_2016, Kohlinger_2017, Hikage_2019}. This small-scale information is heavily influenced by baryonic effects, and the large uncertainty associated with these effects makes them one of the primary sources of systematic error in weak lensing analyses.
Hence, accurate modeling of baryonic effects is crucial when probing the information-rich small scales for an unbiased cosmological analysis. 

At present, numerical simulations remain the sole comprehensive method for precise simulation of baryonic effects and the deeply non-linear evolution of cosmic structures. Hydrodynamic simulations provide a detailed and accurate representation of the behavior of baryonic matter by modeling complex physical processes such as gas dynamics, star formation, and feedback mechanisms \citep{Teyssier_2002, Di_Matteo_2005, Jenkins_1998}. 
However, these simulations are 
not ab initio parameter free, but 
instead must parametrize the lack 
of physics understanding via
free parameters that can be 
varied. 
Furthermore, hydrodynamic simulations require substantial computational resources 
as compared to dark matter only 
N-body simulations. 


The development of emulators has emerged as a powerful technique to overcome this computational challenge, enabling rapid and accurate predictions of physical properties without the need for running costly simulations. 
These emulators interpolate 
simulation results and have been 
shown to be remarkably accurate. 
Various emulators have been developed for cosmology, catering to various observables, encompassing the matter power spectrum \citep{Heitmann_2014, Knabenhans_2019, Winther_2019, Angulo_2021, Euclid_2021}, mass function \citep{McClintock_2019, bocquet2020miratitan}, and the galaxy correlation function and Lyman-$\alpha$ Forest \citep{Zhai_2019, Bird_2019}. 

While most previous work has focused on modeling the baryonic effects on the matter power spectrum \citep{huang2019modelling, mead2021hmcode, arico2021bacco,Schneider_2019,Schneider_2020,giri2023bcemu}, there is an increasing need for developing fast baryon models at the field level for analysis beyond two-point statistics. For example, simulation-based inference methods \citep{Cranmer2020a} 
show great promise in extracting rich non-Gaussian information either through high-order statistics \citep[e.g.,][]{hahn2023simbig}, or directly from the fields \citep[e.g.,][]{Dai2021a,Dai2022a, villaescusa2021multifield, villaescusarobust}. 
These approaches rely on fast and accurate cosmological predictions from numerical simulations.  
Previous field-level baryon models, such as Baryon Correction Model \citep{Schneider_2015} and Enthalpy Gradient Descent \citep{Dai2018a}, move the dark matter particles from N-body simulations to mimic the baryonic effects. While they have been shown to accurately predict the power spectrum from hydrodynamical simulations \citep{Schneider_2019}, they can be computationally expensive when the particle resolution is high.

By analyzing a diverse range of baryonic feedback hydrodynamics simulations across multiple redshifts, we will show that adding baryons to N-body simulations can be achieved using a field-level transfer function to augment N-body fields with a Fourier mode amplitude, k, dependent transfer function correction. We develop a transfer function emulator using Gaussian process for modeling the baryonic effects in terms of $P_{\rm hydro}(k)/P_{\rm nbody}(k)$, where $P_{\rm hydro}(k)$ is the total matter power spectrum, 
and $P_{\rm nbody}(k)$ is the 
dark matter power spectrum. 

In this paper, we develop the 
emulator and show that it is accurate at a percent level over our whole parameter space, which covers scales $0.01 \leq k \leq 10$ h/Mpc and redshifts $0 \leq z \leq 1.5$. We validate the performance of our emulator against thousands of hydrodynamical simulations and their respective gravity-only counterparts. In particular, we make use of CAMELS-Astrid \citep{camels_data_release2}, CAMELS-IllustrisTNG, CAMELS-SIMBA\citep{camels_presentation, camels_data_release1}, 
BAHAMAS \citep{McCarthy_2017, McCarthy_2018}, Horizon AGN \citep{Dubois_2014}, Owls \citep{Schaye_2010, vanDaalen_2011}, and Eagle \citep{Schaye_2015, Crain_2015, McAlpine_2016, Hellwing_2016} simulations.  We also compare against commonly utilized emulators like BACCO \citep{arico2021bacco}, HMcode \citep{mead2021hmcode, HMcode}, and BCemu \citep{giri2023bcemu} on all simulations. Finally, we show the improvement of the field-level baryon model against hydrodynamic fields at varying redshifts. 
This emulator is fast as it only 
requires a single FFT and its inverse, which enables large-volume N-body simulations for generating realistic weak lensing mock data for cosmological analysis at the field level.

This paper is organized as follows: in section \ref{sec:Simulations}, we describe the suite of simulations that are used for training and testing our emulator. In section \ref{sec:field-level} we explain how our emulator can be used to emulate baryonic effects at the field level. In section \ref{sec:GP} we describe the methods and construction of the Gaussian process emulator. In section \ref{sec:Results} we test the emulator's robustness on multiple hydrodynamic test simulations, compare it with currently available emulators, and show field-level improvements using our emulator.  We summarize and conclude in section \ref{sec:Conclusions}.

\section{Simulations}
\label{sec:Simulations}

In this section, we describe the simulations that we employ throughout this paper. Our main suites of simulations are part of the Cosmology and Astrophysics with MachinE Learning Simulations (CAMELS) \citep{camels_presentation, camels_data_release1, camels_data_release2}.
CAMELS is a suite of 10,421 cosmological simulations each with a comoving volume of $(25 ~h^{-1} \text{Mpc})^3$ evolved from $z=127$ to $z=0$ with $256^3$ dark matter particles and $256^3$ gas particles in the initial conditions. These contain 5,097 N-body simulations and 5,324 hydrodynamic simulations. Notably, each hydrodynamic simulation in CAMELS pairs with an N-body counterpart, sharing identical cosmological parameters and initial random seeds. 

Simulations in CAMELS are categorized into various suites (Astrid, IllustrisTNG, and SIMBA) and sets based on the employed code for running the simulations and the arrangement of cosmological and astrophysical parameters ($\Omega_m, \sigma_8, A_{\rm SN1}, A_{\rm AGN}, A_{\rm SN2}, A_{\rm AGN2}$), as well as the initial random seeds. The Astrid suite comprises 1,092 hydrodynamic simulations executed using the MP-Gadget simulation code \citep{yu_feng_2018}, employing analogous subgrid physics as the original Astrid simulations \citep{Ni_2022, Bird_2022}. Additionally, the IllustrisTNG suite \citep[based on][]{Vogelsberger_2013, Torrey_2014} and the SIMBA suite \citep[based on][]{Dave_2016}, with 1,092 hydrodynamic simulations each, are executed using the AREPO code \citep{Springel_2010, Weinberger_2020} and the GIZMO code \citep{Hopkins_2015}, respectively.

Each simulation is characterized by its cosmology (given by $\Omega_m$ and $\sigma_8$) and its astrophysical feedback (given by $A_{\rm SN1}, A_{\rm AGN}, A_{\rm SN2}, A_{\rm AGN2}$). In particular, throughout CAMELS' suites, the astrophysical parameters represent the value of subgrid physics parameters that influence stellar and Active Galactic Nuclei (AGN) feedback mechanisms.

We made use of the Latin Hypercube, LH, set within each suite of CAMELS, which contains 1,000 simulations whose cosmological and astrophysical parameters are arranged in a latin-hypercube within a very broad range\footnote{We note that in the case of Astrid, the parameter $A_{\rm AGN2}$ varies between 0.25 and 4.}
\begin{eqnarray}
\Omega_m &\in& [0.1, 0.5] \\
\sigma_8 &\in& [0.6, 1.0] \\
A_{\text{SN1}}, A_{\rm AGN1} &\in& [0.25, 4.0] \\
A_{\text{SN2}}, A_{\rm AGN2} &\in& [0.5, 2.0]
\end{eqnarray}
and every simulation has a different value of the initial random seed. 
Within each LH set, CAMELS provides 1,000 simulations for each redshift that span a wide range of cosmologies and baryonic feedbacks, perfect for our purposes of capturing the underlying physics.

Importantly, each suite has been run with a different code and therefore the subgrid physics model is completely different. So, notably, while the range of variation of the above parameters remains consistent across all CAMELS suites, the precise definitions and overall impact of these astrophysical parameters vary significantly across suites. These simulations are designed to train and set machine learning algorithms given the way their cosmological, astrophysical, and initial random seed parameters are set. We utilize this set for each suite throughout this paper to train and test our methodology on simulations with significantly different cosmologies and astrophysics. 

\begin{figure*}  
\centering
\begin{multicols}{3}  

\begin{subfigure}{\columnwidth}
  \includegraphics[width=1.1\linewidth]{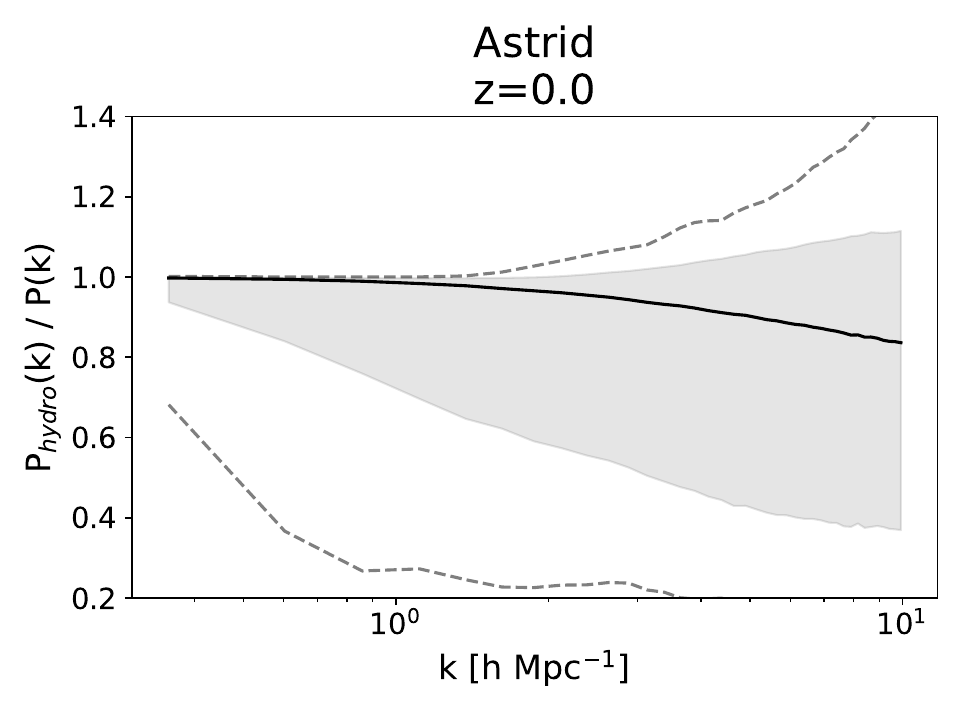}
\end{subfigure}

\begin{subfigure}{\columnwidth}
  \includegraphics[width=1.1\linewidth]{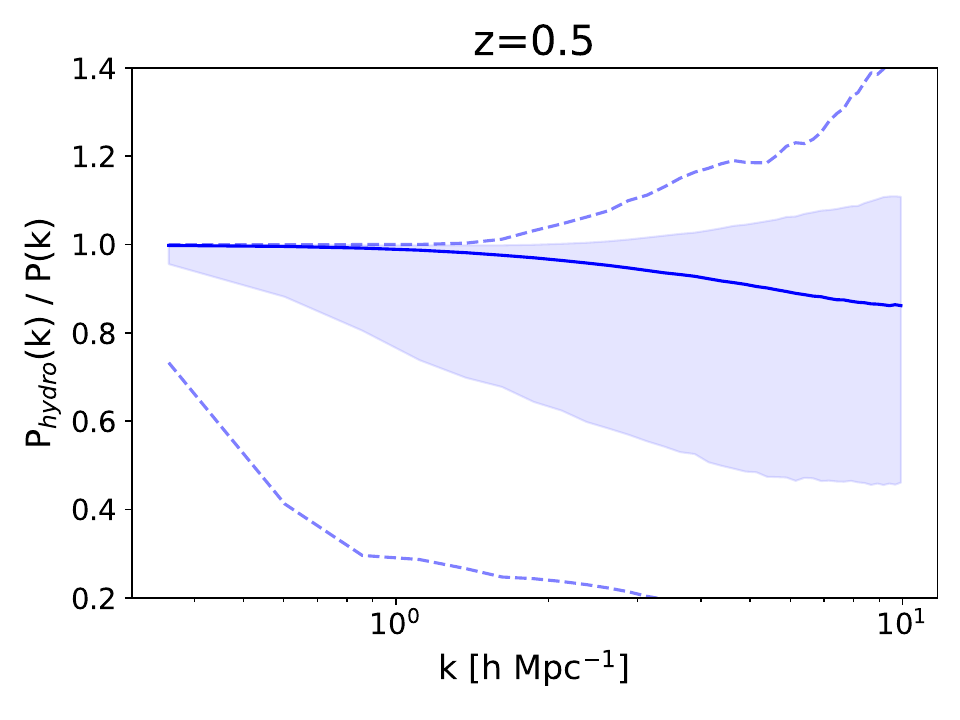}
\end{subfigure}

\begin{subfigure}{\columnwidth}
  \includegraphics[width=1.1\linewidth]{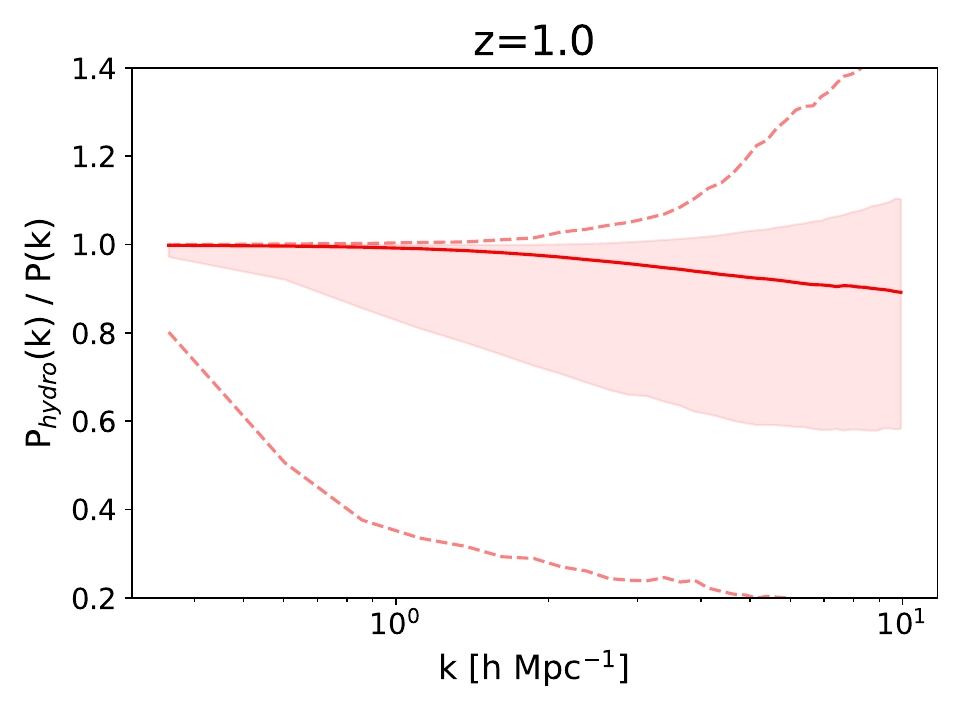}
\end{subfigure}

\begin{subfigure}{\columnwidth}
  \includegraphics[width=1.1\linewidth]{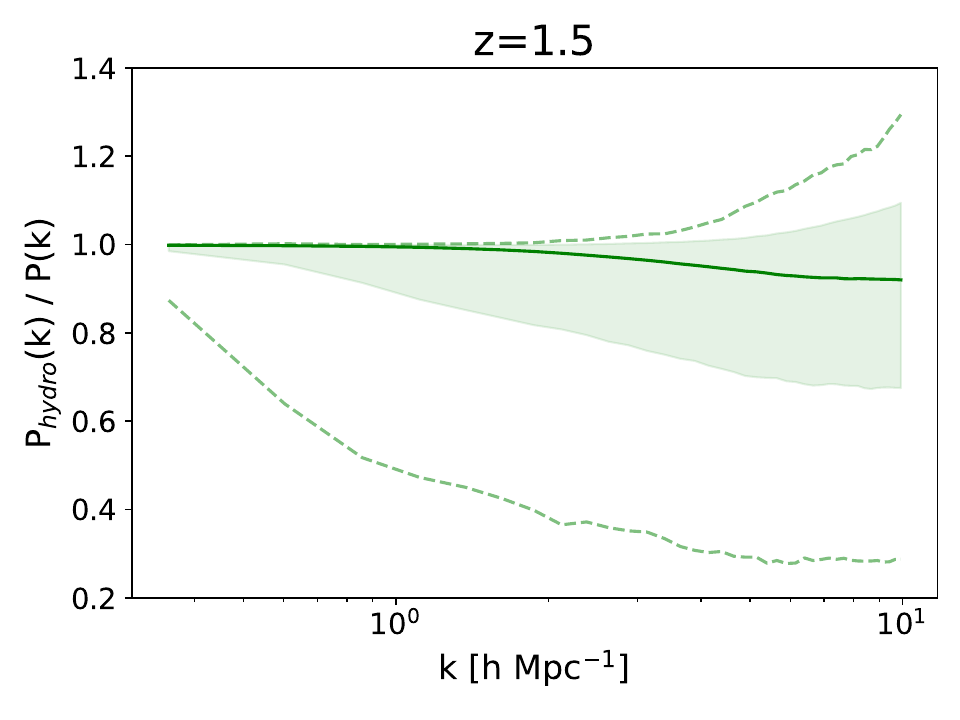}
\end{subfigure}

\columnbreak  

\begin{subfigure}{\columnwidth}
  \includegraphics[width=1.1\linewidth]{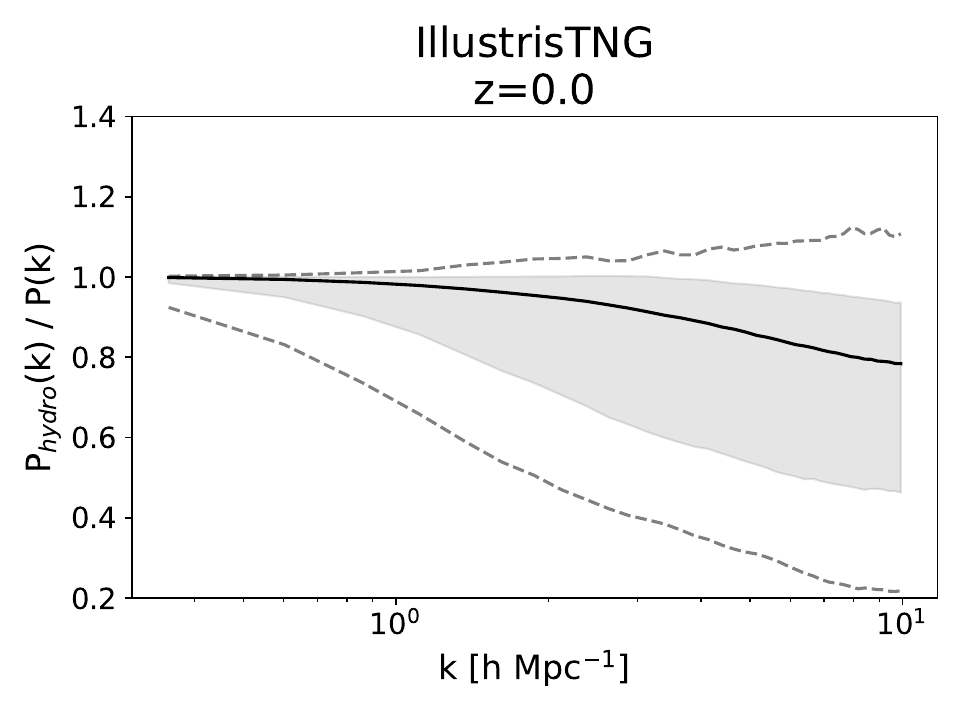}
\end{subfigure}

\begin{subfigure}{\columnwidth}
  \includegraphics[width=1.1\linewidth]{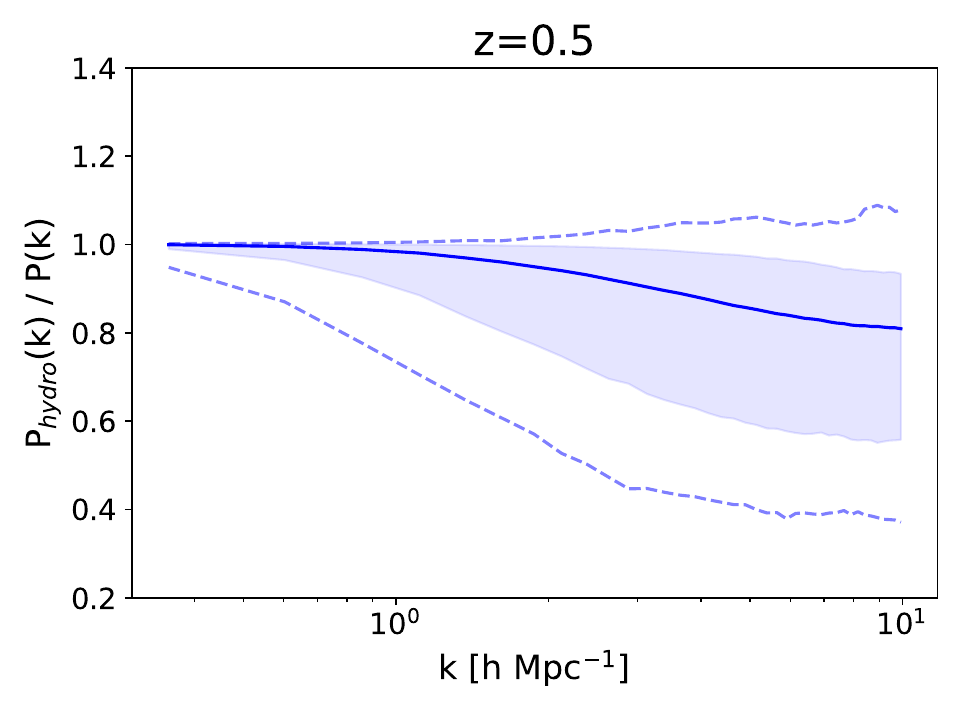}
\end{subfigure}

\begin{subfigure}{\columnwidth}
  \includegraphics[width=1.1\linewidth]{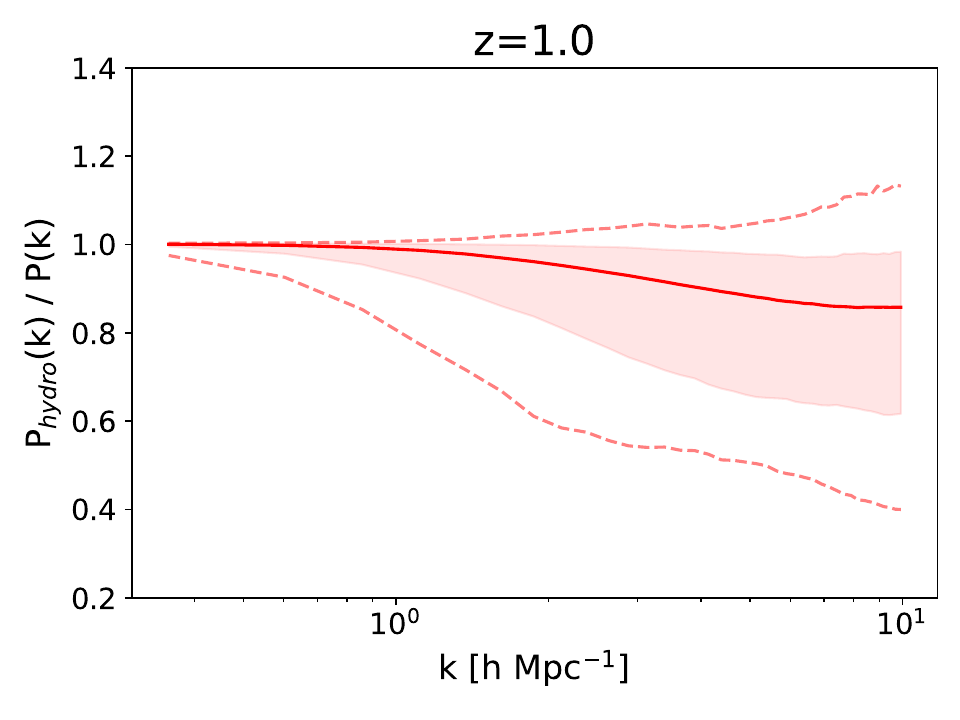}
\end{subfigure}

\begin{subfigure}{\columnwidth}
  \includegraphics[width=1.1\linewidth]{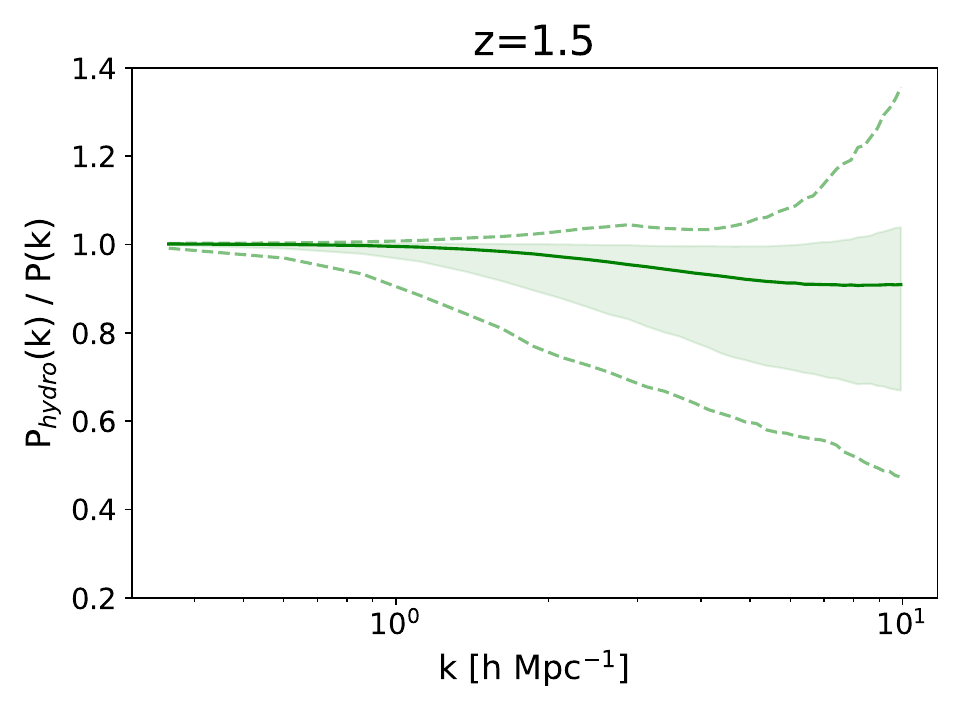}
\end{subfigure}

\columnbreak  

\begin{subfigure}{\columnwidth}
  \includegraphics[width=1.1\linewidth]{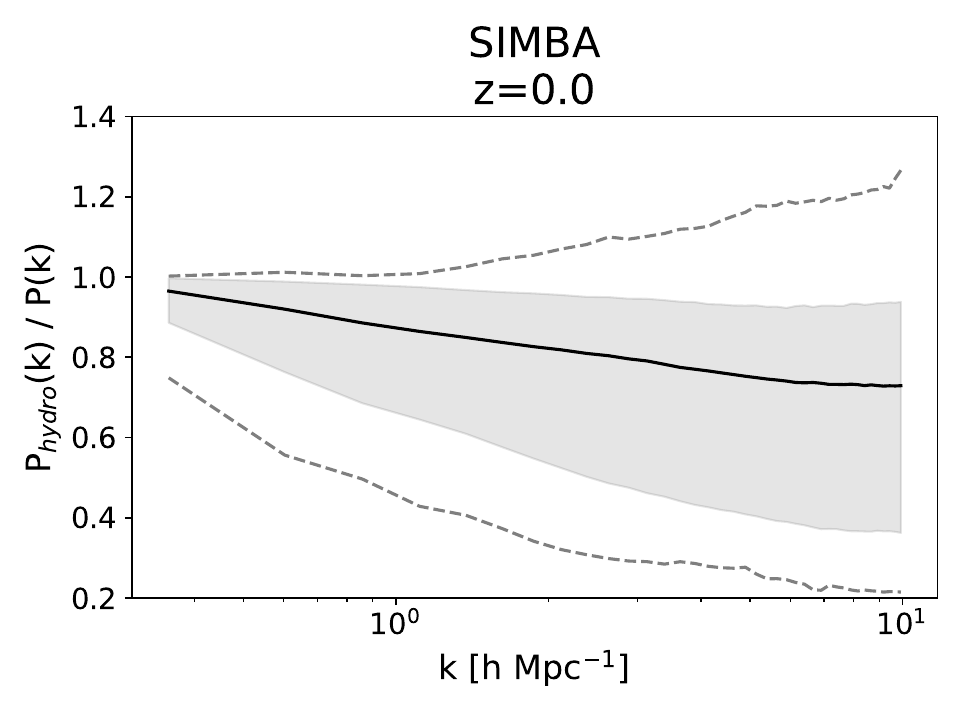}
\end{subfigure}

\begin{subfigure}{\columnwidth}
  \includegraphics[width=1.1\linewidth]{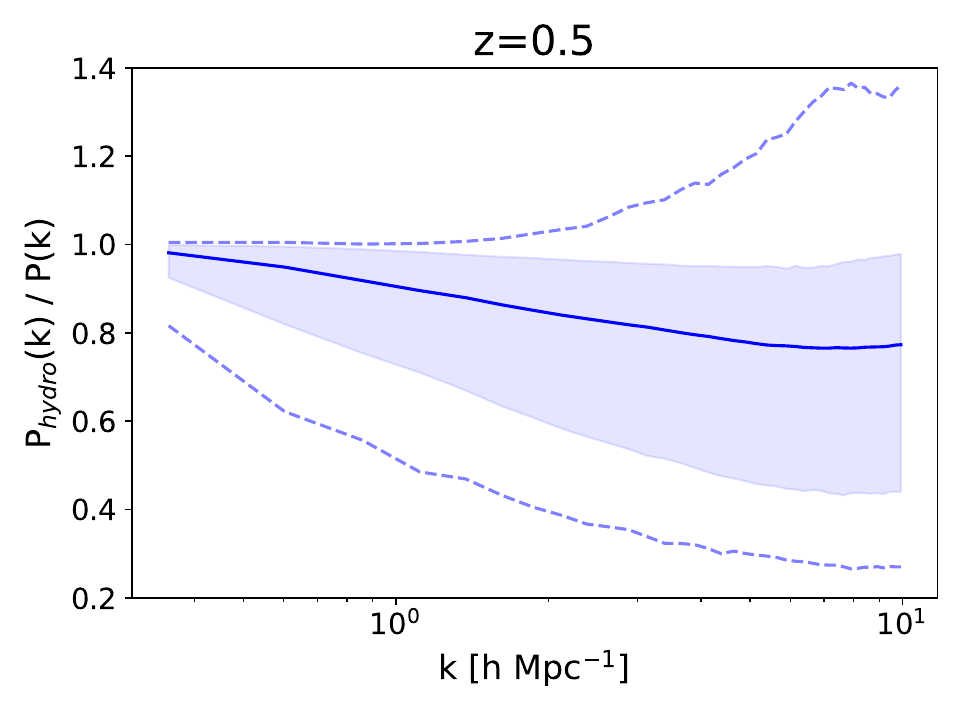}
\end{subfigure}

\begin{subfigure}{\columnwidth}
  \includegraphics[width=1.1\linewidth]{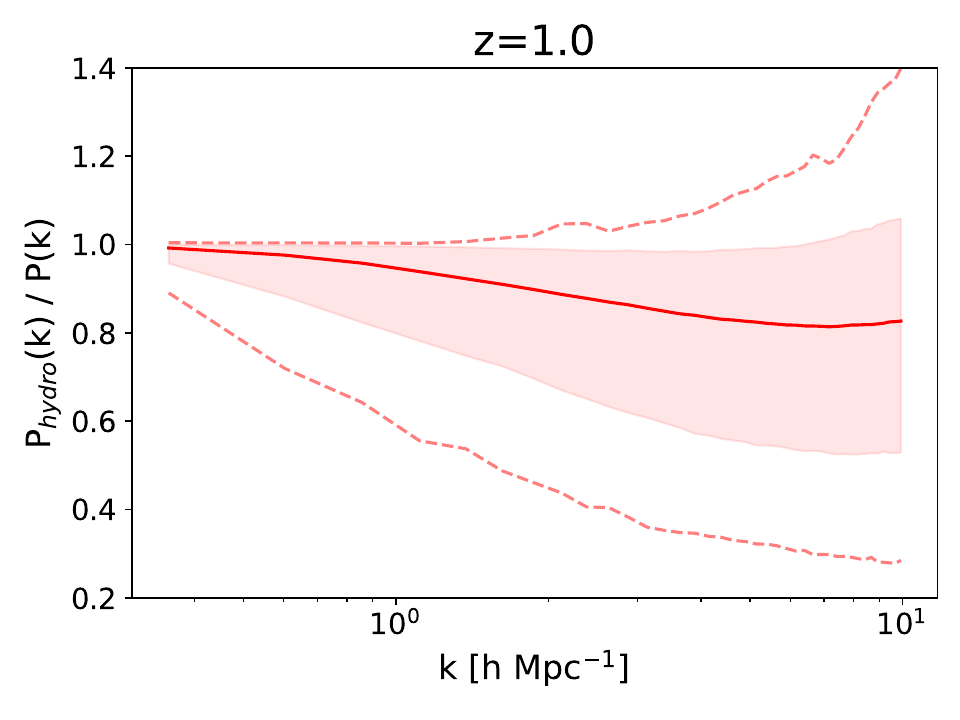}
\end{subfigure}

\begin{subfigure}{\columnwidth}
  \includegraphics[width=1.1\linewidth]{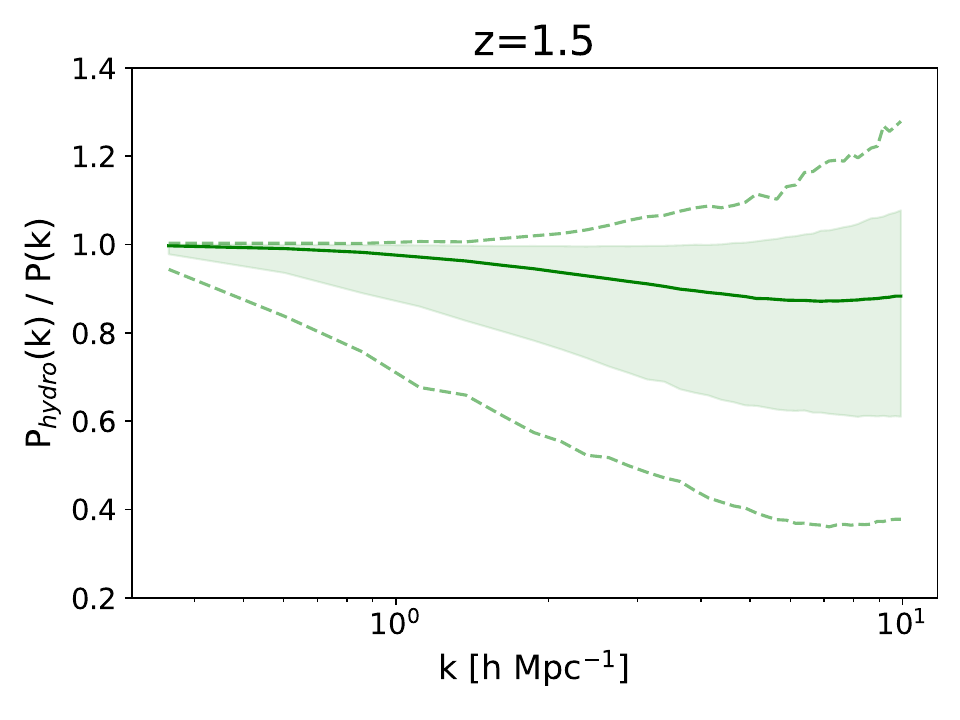}
\end{subfigure}

\end{multicols}
\caption{The matter power spectra ratio observed across suites within CAMELS at various redshifts. Each simulation suite's median value is represented by the solid line, while the dashed lines denote the extreme values, providing an overview of the suite's variability. The shaded region indicates 90 percentiles, reflecting the statistical distribution. During the training phase, our emulator exclusively utilizes 800 Astrid simulations at $z=0.0$. Post-training, we test the emulator on all other CAMELS simulations shown here in addition to the remaining 200 Astrid $z=0.0$ simulations in figure \ref{fig:TestRatios} and other simulations outside CAMELS in figure \ref{fig:OtherHydroTestResults}. Testing the emulator for such varied simulations serves to evaluate the emulator's reliability and generalizability across a wide spectrum of redshifts and simulations.}
\label{fig:CAMELSRatios}
\end{figure*}

Figure \ref{fig:CAMELSRatios} shows the baryonic effect
in the matter power spectrum, $P_{\rm hydro}(k)/P_{\rm nbody}(k)$, across different redshift values in all CAMELS suites utilized in this study. From the figure, it is clear that baryonic feedback can have very diverse and strong effects on the matter power spectrum, especially on small scales. SIMBA, with its aggressive AGN feedback, produces the most prominent suppression of the matter power on large scales. On the other hand, IllustrisTNG exhibits a more moderate impact on the matter power spectrum as a consequence of having milder AGN feedback, and Astrid spans the widest range of baryonic feedback, encompassing effects seen in both SIMBA and IllustrisTNG. 

To cover the broadest range of baryonic feedback, based on Figure \ref{fig:CAMELSRatios}, we used simulations from the Astrid suite at $z=0.0$ to train our emulator. For selecting the training simulations, we performed a random sampling of 800 simulations (out of the 1000 available) from the Astrid suite within CAMELS at $z=0.0$.

\begin{figure}
\centering
\includegraphics[width=1.0\linewidth]{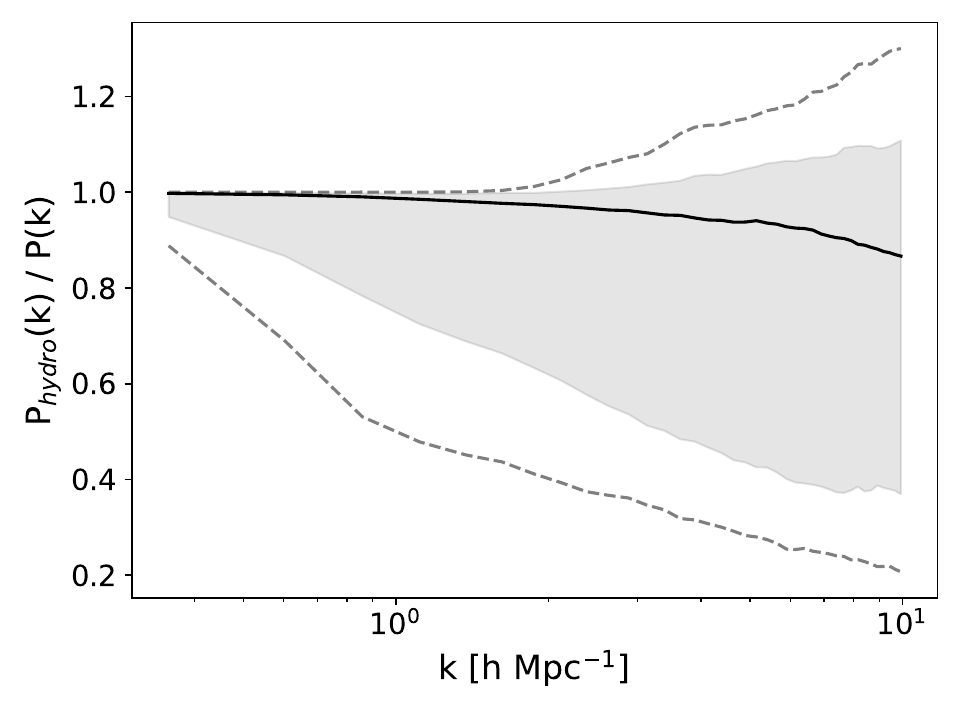} 
\caption{The ratio of matter power spectra in the 200 Astrid test simulations for $z=0.0$. The solid line denotes the median value across the simulations, while the dashed lines illustrate the extreme values. The shaded region delineates the 90th percentile range. During emulator training, 800 simulations were randomly sampled from the Astrid (at $z=0.0$) suite within CAMELS, leaving these 200 remaining simulations for post-training Astrid $z=0.0$ test analysis.}
\label{fig:TestRatios}
\end{figure}

Post-training, we test the emulator using the remaining 200 simulations from Astrid at $z=0.0$ alongside all other available hydrodynamic simulation suites in CAMELS. Figure \ref{fig:TestRatios} illustrates the matter power spectrum ratio in the Astrid $z=0.0$ test data. 
Additionally, the IllustrisTNG suite and the SIMBA suite are part of the test dataset. 

\begin{figure}
\centering
\includegraphics[width=1.0\linewidth]{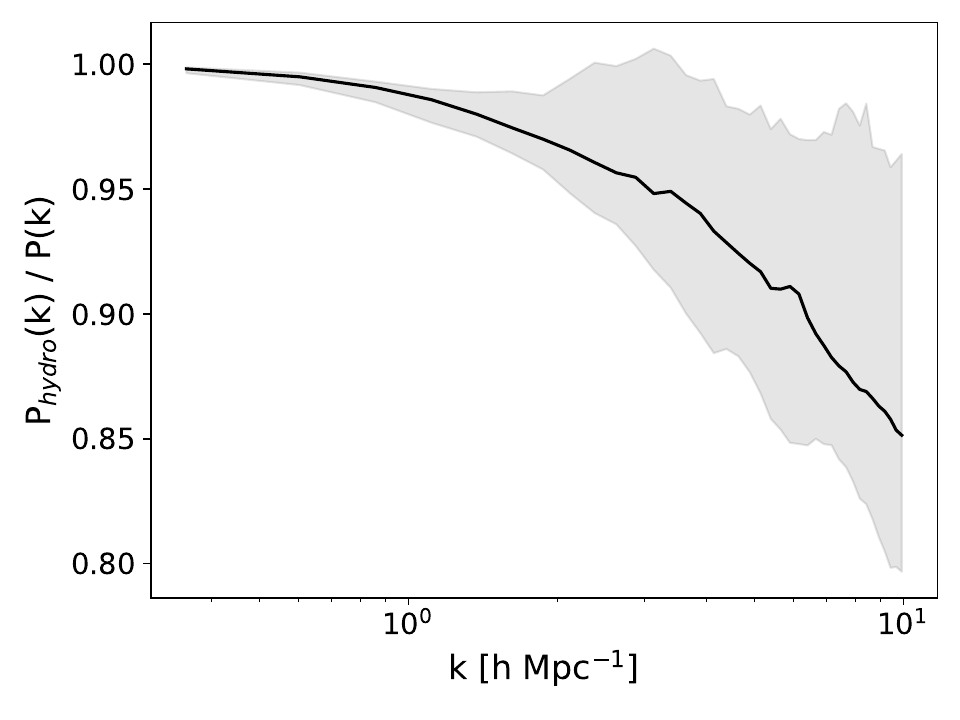} 
\caption{Ratios of the matter power spectra in the Astrid $z=0.0$ CV set. The solid line shows the median value, while the shaded region shows the entire range of the ratios. All the simulations share the value of the cosmological and astrophysical parameters; they only differ in the value of their initial conditions random seed and hence can be used to study the effect of cosmic variance.}
\label{fig:CVratios}
\end{figure}

Previous studies \citep{Heitmann_2014, Smith_2014, Rasera_2014, Heitmann_2013} have shown that both high physical resolution and large box sizes are required to guarantee convergence of the power spectrum. \citet{Schneider_2015} showed that deviations of the power spectrum ratio using small boxes, like in CAMELS, are at the $5\%$ level. However, these could also be due to cosmic variance which affects the small CAMELS-like box volumes. To study the effects of cosmic variance on the matter power spectrum ratio, we show the ratios for simulations with the same cosmology and astrophysics from CAMELS-Astrid's CV set in figure \ref{fig:CVratios}. From this, we can see that the matter power spectrum ratio is affected by cosmic variance up to $\sim 10\%$ on small scales, suggesting that the deviations for small boxes are due to cosmic variance.

In addition to the diverse array of simulations in CAMELS, we extended the validation of our emulator by testing it against simulations outside the CAMELS database. These external simulations, including BAHAMAS \citep{McCarthy_2017, McCarthy_2018}, Horizon AGN \citep{Dubois_2014}, Owls \citep{Schaye_2010, vanDaalen_2011}, and Eagle \citep{Schaye_2015, Crain_2015, McAlpine_2016, Hellwing_2016}, serve as crucial benchmarks to assess the robustness and generalizability of the emulator's predictions beyond CAMELS. 
These simulations encompass various diverse physical processes including AGN feedback, supernovae feedback, mass loss from Asymptotic Giant Branch stars, radiative cooling, stellar winds, and stellar initial mass function, among others. Additionally, these simulations differ from those in CAMELS in volume, resolution, and subgrid physics code \citep{vanDaalen_2020}, enabling us to test the robustness of our emulator to these effects. 
The solid lines in the left panel of Figure \ref{fig:OtherHydroTestResults} illustrate the matter power spectrum ratio derived from these external simulations, allowing us to assess how the emulator performs across varied simulations beyond the scope of the CAMELS.


\section{Baryonic effects at the field-level}\label{sec:field-level}
In this section, we present our methodology to model baryonic effects at the field level for the total matter density field. 
We start by computing the cross-correlation coefficient between the matter field in full hydrodynamic simulations and their N-body counterparts. The cross-correlation coefficient, $r(k)$, is defined as:
\begin{align}
r(k) = \frac{P_{\rm cross}(k)}{\sqrt{P_{\rm nbody}(k)P_{\rm hydro}(k)}}
\end{align}

Here, $P_{\rm cross}(k)$, $P_{\rm nbody}(k)$, and $P_{\rm hydro}(k)$ represent the cross-power spectrum, the N-body power spectrum, and the power spectrum of the hydrodynamic fields respectively. This coefficient's range spans from $-1$ to $1$, where values closer to $1$ signify a strong positive linear relationship, $-1$ indicates a strong negative linear relationship and $0$ implies no linear relationship between the datasets.

Figure \ref{fig:CAMELS_CC} shows the cross-correlation coefficients derived from all CAMELS suites at different redshift values pertinent to this study. These coefficients serve as indicators of the correlation strength between N-body and hydrodynamic fields within the simulations.

We can see that the calculated cross-correlation coefficients are very close to 1, down to $k \sim 10$ h/Mpc, for all the simulations, with most deviations being within $5-10\%$. On the other hand, we see in Figure \ref{fig:CAMELSRatios} that baryonic effects can cause deviations of up to $\sim 50\%$ on the matter power spectrum, with the effects getting more dominant at smaller scales. This suggests that the baryonic effects predominantly impact the amplitude of the Fourier modes (given by the power spectrum) rather than their phases (given by cross-correlation coefficients).

Since the amplitude and phase of the Fourier modes completely describe the fields, with baryonic effects mainly changing the amplitudes, aligning the power spectra of N-body fields with their hydrodynamic counterparts would also effectively align them at the field level, facilitating a cost-effective field-level analysis using just N-body simulations.

We can achieve this power spectra alignment by applying a transfer function to the N-body fields \citep{Bond_1983, Seljak_Zaldarriaga_1996}. Transfer functions operate by performing specific modifications to the field data. In Fourier space, each mode of a field is represented by its amplitude, typically denoted by $|\mathbf{k}|$, and its phase. Transfer functions act on $\mathbf{k}$ to alter their amplitudes according to certain criteria \citep{Peacock_1999, Dodelson_2003}.

The field-level transformation using a transfer function, $T(k)$, is mathematically defined as:
\begin{align}
F'(\mathbf{k}) = T(k) \cdot F(\mathbf{k})
\end{align}
Here, $F(\mathbf{k})$ symbolizes the original field in Fourier space, and $F'(\mathbf{k})$ represents the transformed field in Fourier space after the element-wise application of the transfer function $T(k)$ to the original field $F(\mathbf{k})$. This transformation enables the adjustment of simulated fields to match desired characteristics or observational data, enhancing the accuracy or realism of the simulation results.

In our context of incorporating baryonic effects in N-body simulations, we apply a transfer function to the simulated field to adjust its power spectrum. Defining baryonic suppression as:
\begin{align}
S(k) \coloneqq P_{\rm hydro}(k)/P_{\rm nbody}(k)
\end{align}
Our field transformation on N-body fields is then:
\begin{align}
\delta_{\rm nbody}'(\mathbf{k}) = \sqrt{S(k)} \cdot \delta_{\rm nbody}(\mathbf{k}) \label{eq:field-transformation}
\end{align}
This could increase or suppress the power of certain scales and correct discrepancies arising from missing baryonic physics effects in N-body simulations, aligning the power spectra of N-body fields with the full hydrodynamic fields.

\begin{figure*}  
\centering
\begin{multicols}{3} 
\begin{subfigure}{\columnwidth}
  \includegraphics[width=1.1\linewidth]{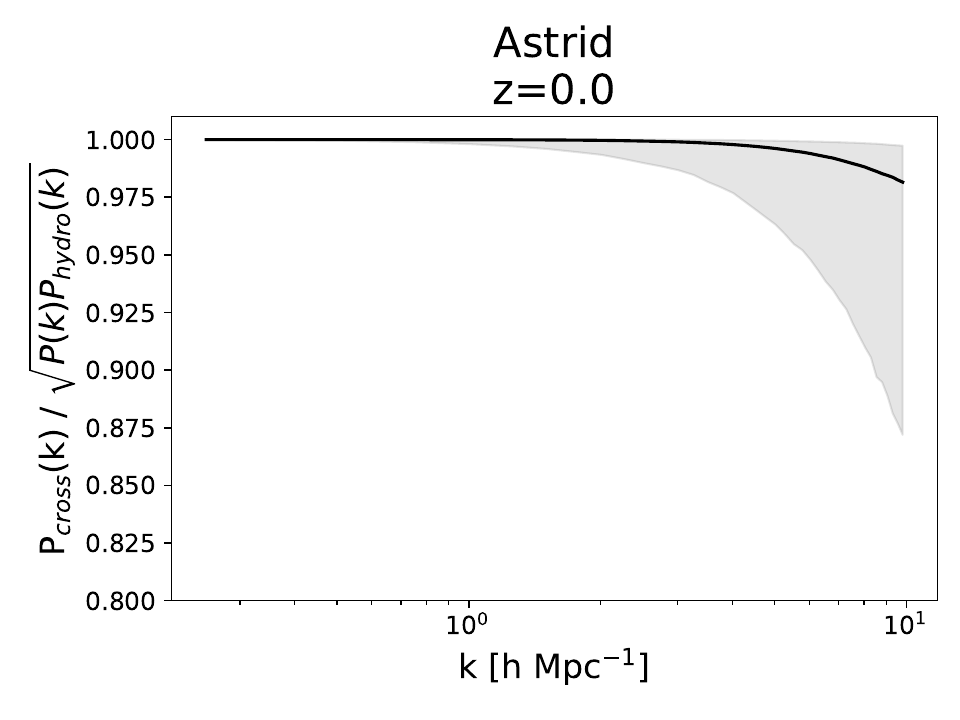}
\end{subfigure}

\begin{subfigure}{\columnwidth}
  \includegraphics[width=1.1\linewidth]{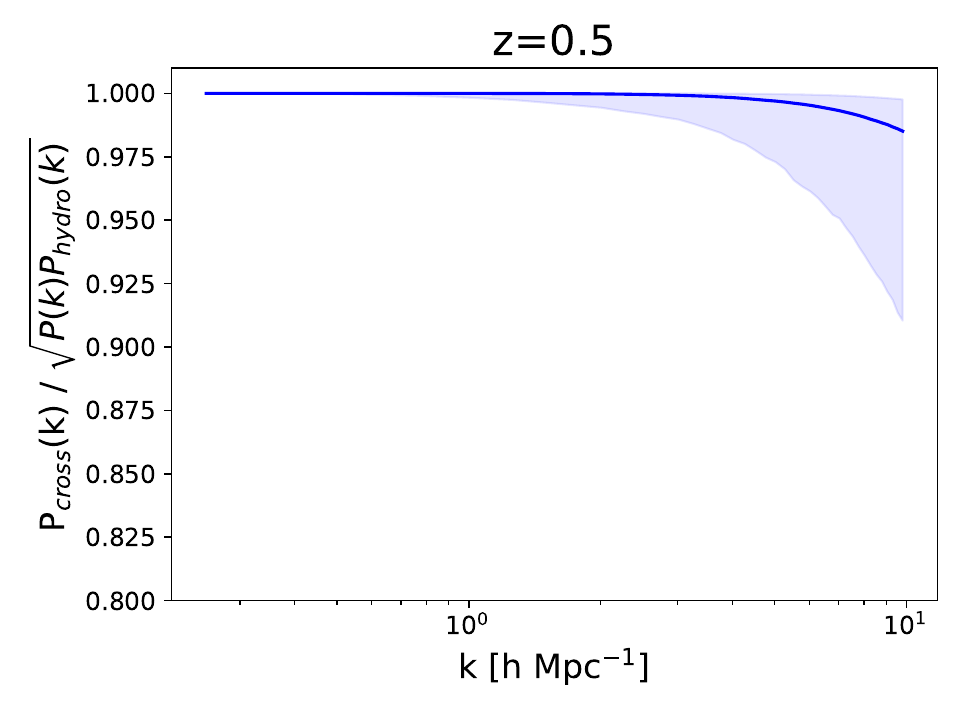}
\end{subfigure}

\begin{subfigure}{\columnwidth}
  \includegraphics[width=1.1\linewidth]{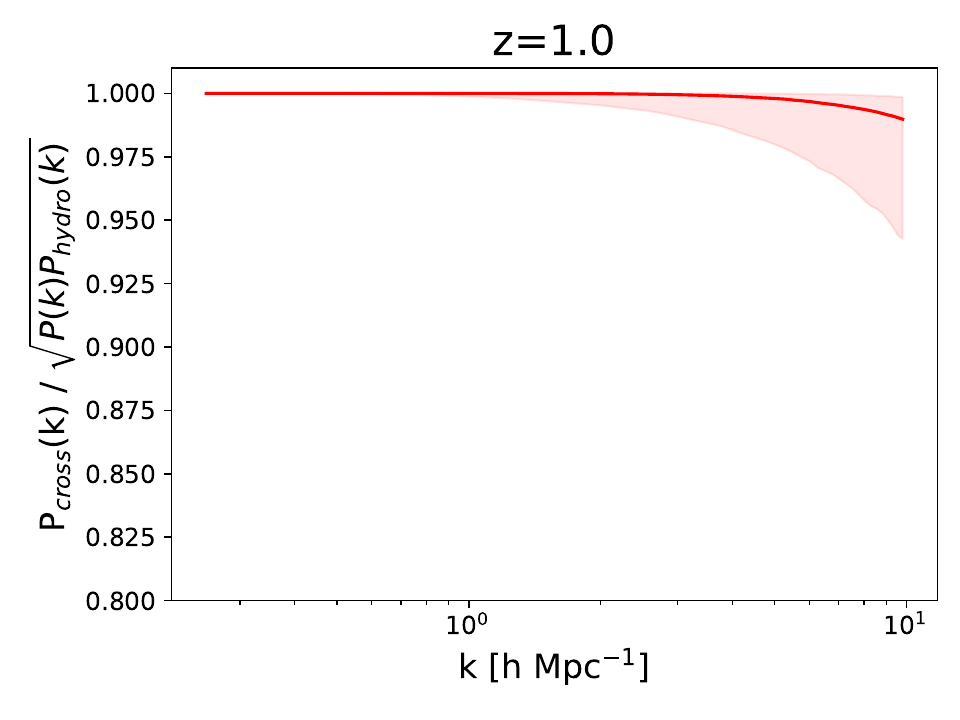}
\end{subfigure}

\begin{subfigure}{\columnwidth}
  \includegraphics[width=1.1\linewidth]{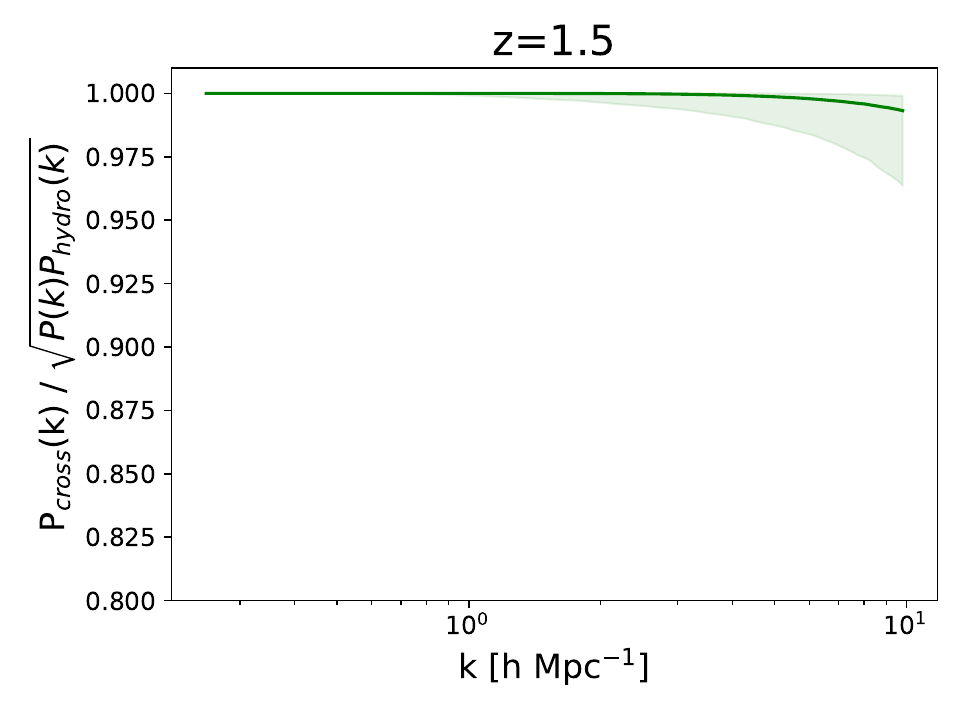}
\end{subfigure}

\columnbreak  

\begin{subfigure}{\columnwidth}
  \includegraphics[width=1.1\linewidth]{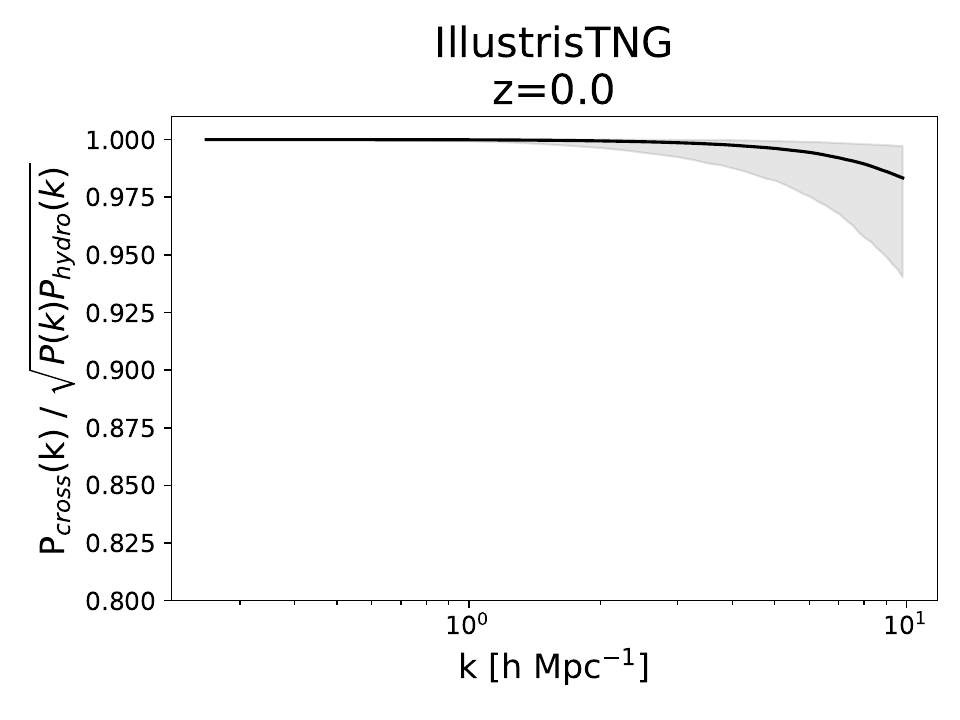}
\end{subfigure}

\begin{subfigure}{\columnwidth}
  \includegraphics[width=1.1\linewidth]{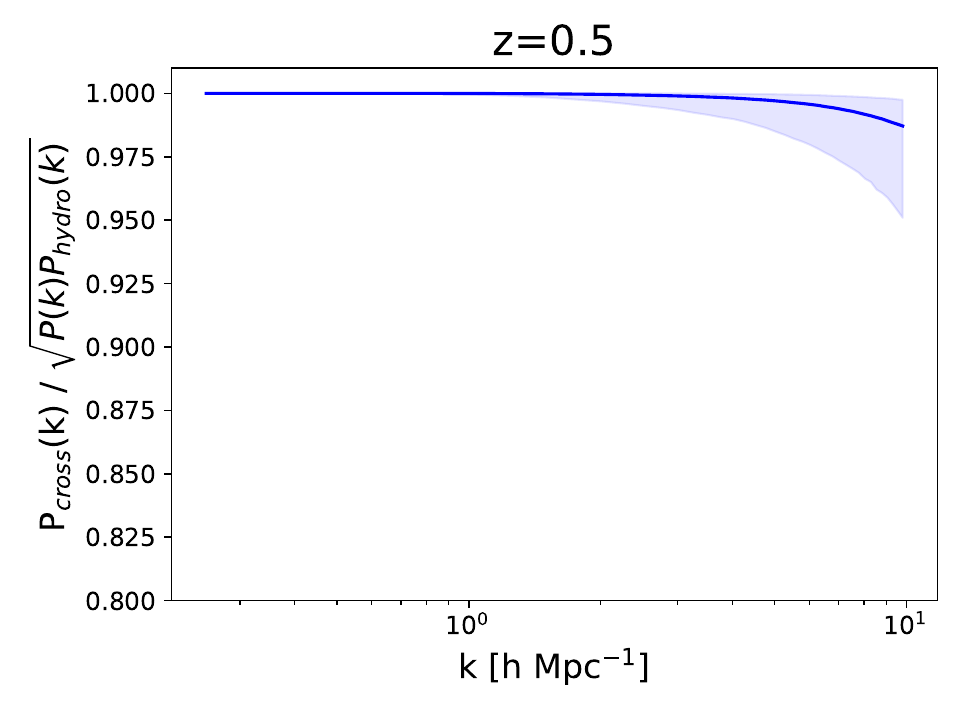}
\end{subfigure}

\begin{subfigure}{\columnwidth}
  \includegraphics[width=1.1\linewidth]{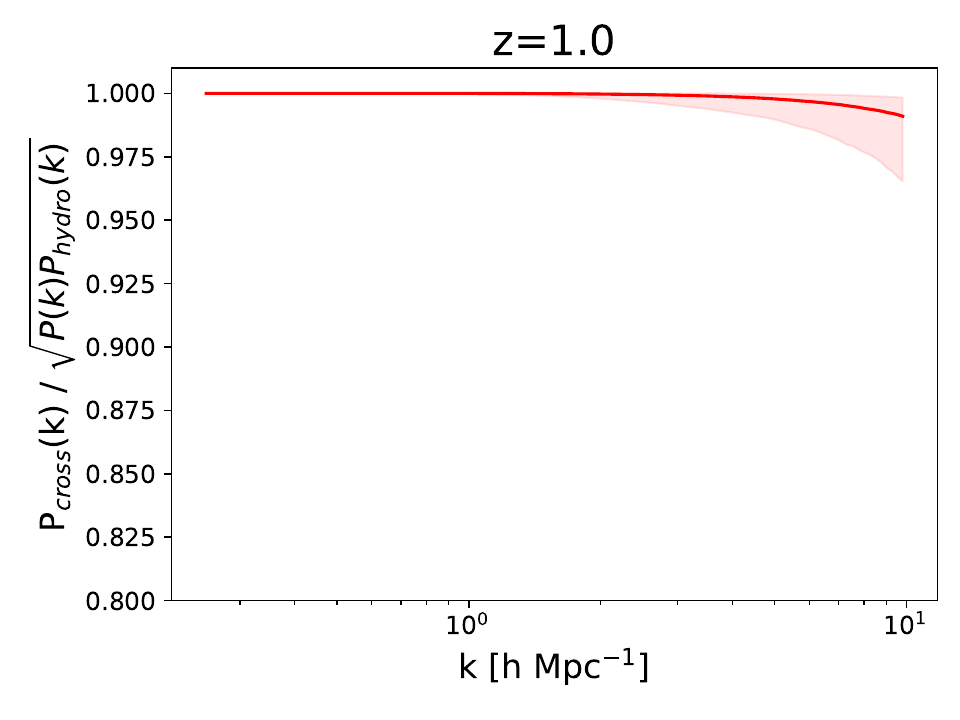}
\end{subfigure}

\begin{subfigure}{\columnwidth}
  \includegraphics[width=1.1\linewidth]{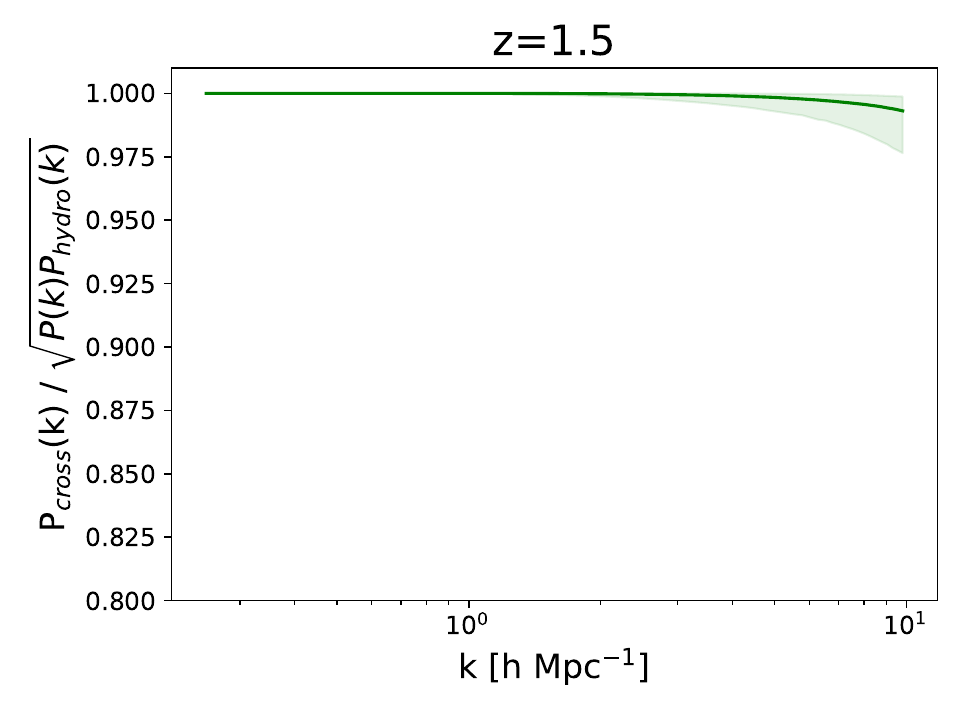}
\end{subfigure}

\columnbreak  

\begin{subfigure}{\columnwidth}
  \includegraphics[width=1.1\linewidth]{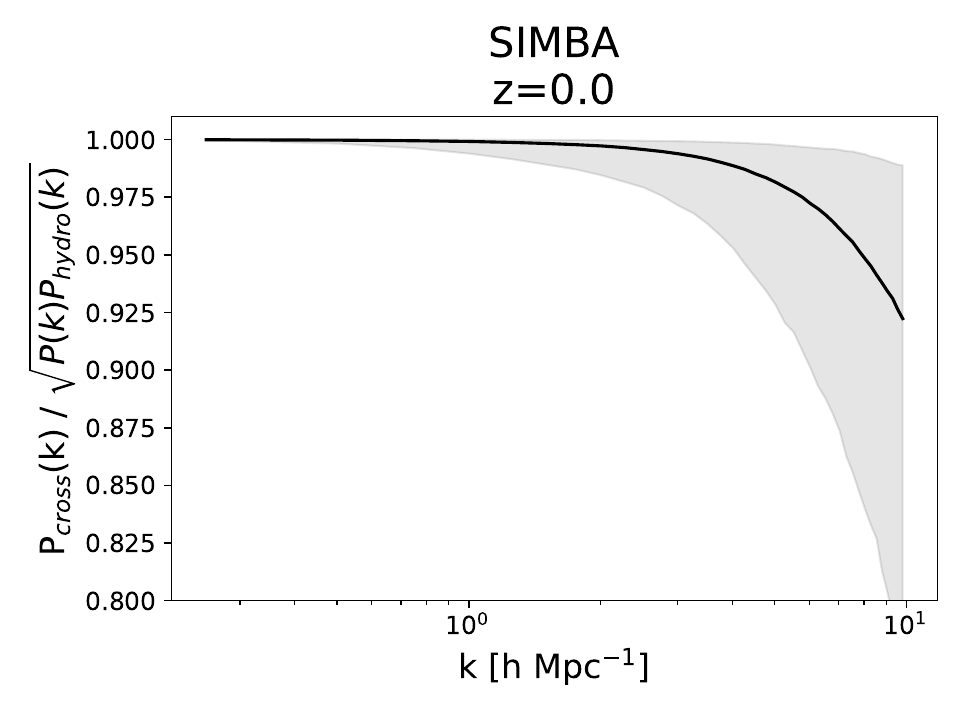}
\end{subfigure}

\begin{subfigure}{\columnwidth}
  \includegraphics[width=1.1\linewidth]{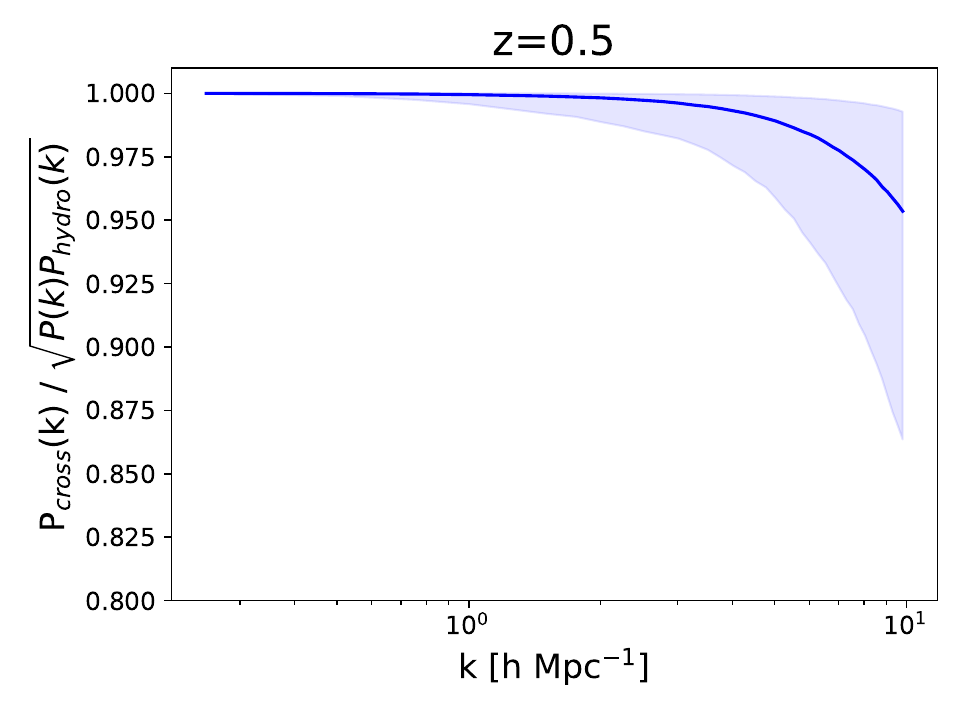}
\end{subfigure}

\begin{subfigure}{\columnwidth}
  \includegraphics[width=1.1\linewidth]{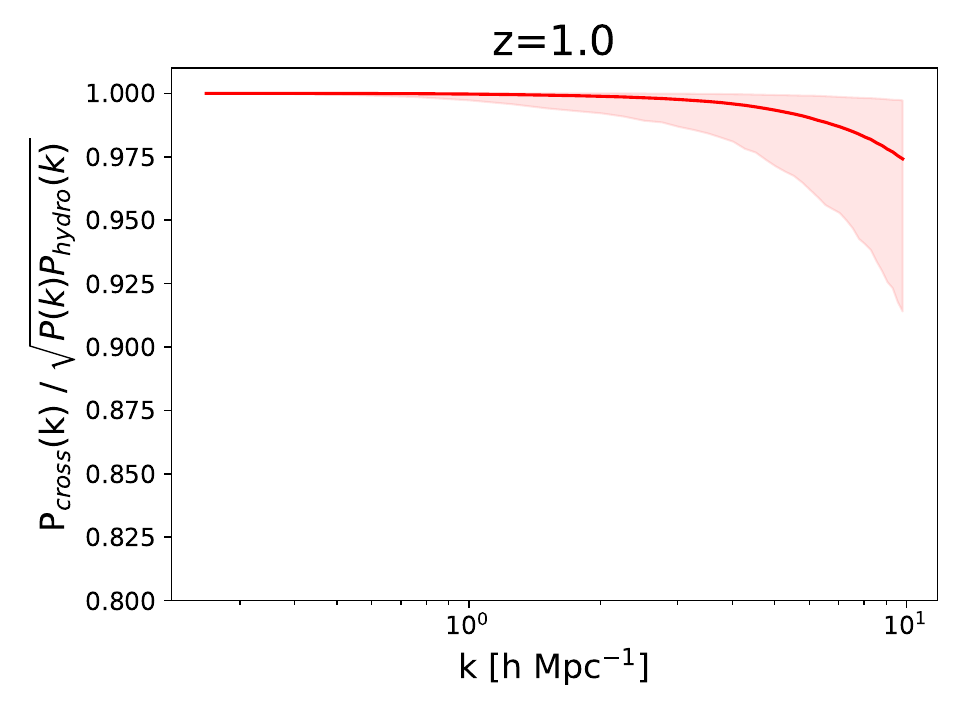}
\end{subfigure}

\begin{subfigure}{\columnwidth}
  \includegraphics[width=1.1\linewidth]{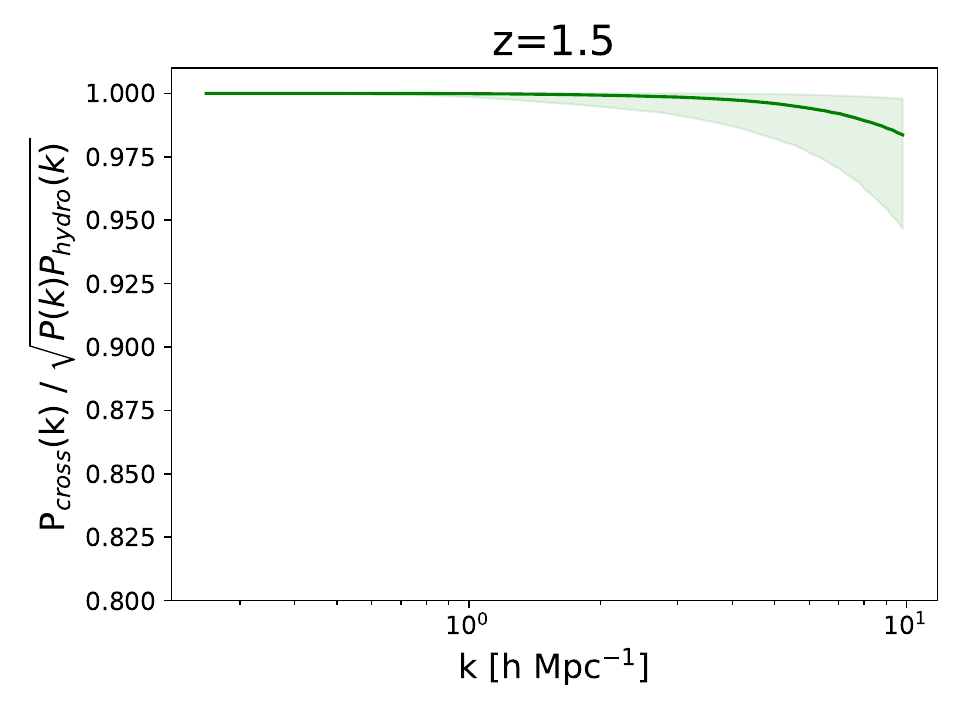}
\end{subfigure}
\end{multicols}
\caption{The cross-correlation coefficients between N-body and hydrodynamic fields across all suites within CAMELS at different redshifts. The solid line signifies the median value, while the shaded region delineates the 90th percentile range, reflecting the distribution and variability of these coefficients. Remarkably, the computed cross-correlation coefficients for all simulations are very close to 1, denoting a strong relationship between N-body and hydrodynamic fields. This suggests that the impact of baryonic effects predominantly alters the amplitude (power) of the Fourier modes while exerting minimal influence on their phase (cross-correlation coefficient). Consequently, we posit that field-level enhancements in N-body fields can be achieved by refining their power spectra to closely align with the corresponding hydrodynamic fields. Targeting the power spectra improvement holds promise for effectively reconciling the discrepancies between N-body and hydrodynamic simulations, facilitating more accurate and cost-effective field-level analyses across simulations.}
\label{fig:CAMELS_CC}
\end{figure*}
\label{fig:CAMELS_CC}

\section{Gaussian Process Emulator}\label{sec:GP}
To fulfill the promise of field-level modeling of baryonic effects, we need to characterize $P_{\rm hydro}(k)/P_{\rm nbody}(k)$ for our transfer function. 
In this section, we describe the main numerical methods we use to create our emulator of $P_{\rm hydro}(k)/P_{\rm nbody}(k)$ using Gaussian Processes. 

Gaussian Processes (GPs) \citep{Rasmussen_2004} are a versatile tool within machine learning and statistics, renowned for their efficacy in regression, interpolation, and uncertainty quantification. They provide a flexible framework for modeling functions along with their associated uncertainties \citep[e.g.][]{Bird_2019, Rogers_2019, Rogers_2021, Pedersen_2021}.
Moreover, a Gaussian process emulator is computationally efficient, enabling its use within standard inference methodologies such as Markov Chain Monte Carlo (MCMC) for evaluations.

For the length scales we want to model, the CAMELS simulations have 39 linearly-spaced $k$ bins spanning the range $0.36 < k < 9.93$ h/Mpc. Hence, to model baryonic effects at small scales down to $k \sim 10$, our emulator delineates these 39 linearly-spaced $k$ bins. We treat the baryonic effects in each $k$ bin as an individual Gaussian process, enabling independent training for each bin, with every simulation serving as a training point for these $k$ bins.

At a specific point $X = [\Omega_m, \sigma_8, A_{\rm SN1}, A_{\rm AGN}, A_{\rm SN2}, A_{\rm AGN2}]$ in the parameter space, a Gaussian process models the target function — $S(\mathbf{k}|X) \coloneqq P_{\rm hydro}(\mathbf{k}|X)/P_{\rm nbody}(\mathbf{k}|X)$ in our case — as an assembly of random variables that form a joint Gaussian distribution. This model is defined via $S(\mathbf{k}|X) \sim \mathcal{N}(0, K(X, X_i))$, where $X_i$ signifies the parameter values at the training simulations, and $K(X, X')$ represents a covariance kernel.

The choice of the kernel function $K(X, X_i)$ plays a pivotal role in characterizing the correlation or similarity between data points $X$ and $X_i$. This function serves as a prior, encapsulating the expected behavior of the underlying function — baryonic effects in our case — to be modeled. For our emulator, we adopt a Matérn 5/2 kernel, a generalized form of the radial basis function (RBF), defined by:

\begin{align}
K(X, X') = \sigma_{0}^2 \left(\sum_{i=1}^{6} \left(1 + \frac{\sqrt{5}r}{\ell_i} + \frac{5r^2}{3\ell_i^2}\right) \exp\left(-\frac{\sqrt{5}r}{\ell_i}\right)\right)
\end{align}

Here, $r = || X - X' ||_2$ represents the L2 distance between two data points, $\sigma_{0}^2$ denotes the variance parameter, and $\ell_i$ signifies the length scale of each input dimension, influencing the smoothness and range of correlations between data points. Our choice of this covariance kernel is motivated by the need for flexibility, achieved through the squared exponential kernel, the efficacy of linear interpolation, and allowing for noise in the training data.

At a test point $X_{*}$, the joint distribution of the test data $S(\mathbf{k}|X_*)$ and training data $S(\mathbf{k}|X_i)$ can be expressed as:

\begin{align}
\begin{bmatrix}
S(\mathbf{k}|X_i) \\
S(\mathbf{k}|X_*)
\end{bmatrix}
\sim \mathcal{N}\left(0, \begin{bmatrix}
K(X_i, X_i) + \sigma_n^2 I & K(X_i, X_*) \\
K(X_*, X_i) & K(X_*, X_*)
\end{bmatrix}\right)
\end{align}

Here, $\sigma_n^2$ serves as a hyperparameter signifying Gaussian noise within the training data.

Consequently, our Gaussian process involves a total of 8 hyperparameters: 6 correlation lengths, $\sigma_0^2$, and $\sigma_n^2$. The hyperparameters are optimized by maximizing the marginal log-likelihood of the training data \citep{Rasmussen_2004}.

The posterior predictive distribution over test data is obtained through:

\begin{align}
S(\mathbf{k}|X_*) \sim \mathcal{N}(\mu, \Sigma)\\
\mu = K(X_*, X_i) \left(K(X_i, X_i) + \sigma_n^2 I\right)^{-1} S(\mathbf{k}|X_i),\\
\Sigma = K(X_*, X_*) - K(X_*, X_i) \left(K(X_i, X_i) + \sigma_n^2 I\right)^{-1} K(X_i, X_*)
\end{align}

The mean and variance derived from the posterior predictive distribution, using the training information at $X_i$, serve as estimators for the value and interpolation uncertainty associated with $S(X_*)$.

We implement our emulator using tinygp \citep{dan_foreman_mackey_2022}, a Python library for GP Regression (GPR) built on top of the JAX library for numerical computing \citep{bradbury2018jax}. 

The GP model offers a broad prior across function space, enabling the modeling of the diverse baryonic effects we see in figure \ref{fig:CAMELSRatios} without imposing strong prior constraints on its parameter dependencies. Since it is stochastic, this model provides predictions for the baryonic suppression beyond the training points, accompanied by associated uncertainties that can be integrated into our statistical model.
This emulation methodology provides a robust approach for modeling and predicting the baryonic effects in simulations, enabling efficient and accurate interpolation, and quantification of uncertainties.

\section{Results}
\label{sec:Results}

In this section, we use the trained Gaussian Process emulator to generate predictions for $P_{\rm hydro}(k)/P_{\rm nbody}(k)$ as a function of four astrophysical parameters - $A_{\rm SN1}, A_{\rm AGN}, A_{\rm SN2}, A_{\rm AGN2}$ within their respective ranges. We emphasize that our emulator was trained on Astrid simulations at $z=0$, and therefore, the meaning of these astrophysical parameters is, in principle, associated with the Astrid subgrid physics model. However, to make our emulator generic and robust, from now on, we will consider these four astrophysical parameters as nuisance parameters that one needs to tune to reproduce the result of one particular hydrodynamic simulation.


We employ the differential evolution global optimizer from the SciPy library \citep{Virtanen_2020} to obtain the best-fit value of these nuisance parameters. This optimization technique is adept at exploring the parameter space to seek optimal solutions, especially in scenarios with complex, multi-dimensional parameter spaces. The differential evolution \citep{Storn_1996, Storn_1997, Price_2005} method operates stochastically, offering a non-gradient approach to locating the minimum and can search through large volumes in parameter space.

We now show the accuracy of our emulator for simulations within and outside CAMELS, showing its precision to changes in simulation cosmology, feedback, subgrid physics, resolution, volume, and redshift. On top of that, we compare the accuracy of our emulator against other emulators in the literature. Finally, we demonstrate the emulator's efficacy in creating field-level improvements when applied to the N-body fields of simulations within CAMELS, validating its potential for advancing field-level weak-lensing analyses using cosmological simulations.


\subsection{Emulator accuracy}

We start by quantifying the accuracy of our emulator across hydrodynamic simulations.

\begin{figure*}
\centering
\begin{multicols}{3} 

\begin{subfigure}{\columnwidth}
  \includegraphics[width=1.1\linewidth]{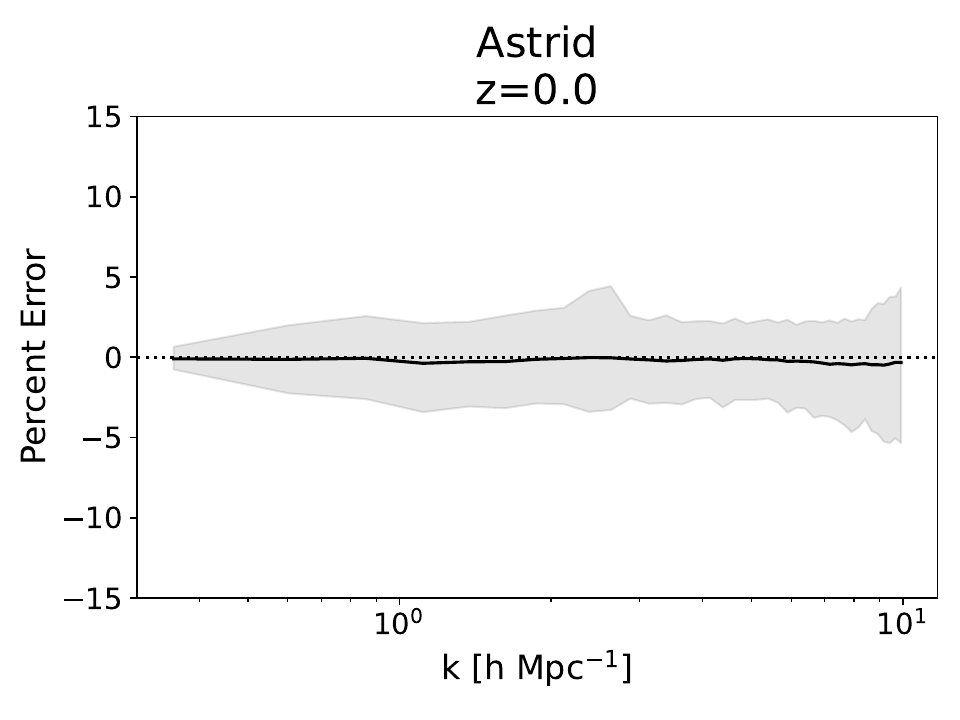}
\end{subfigure}

\begin{subfigure}{\columnwidth}
  \includegraphics[width=1.1\linewidth]{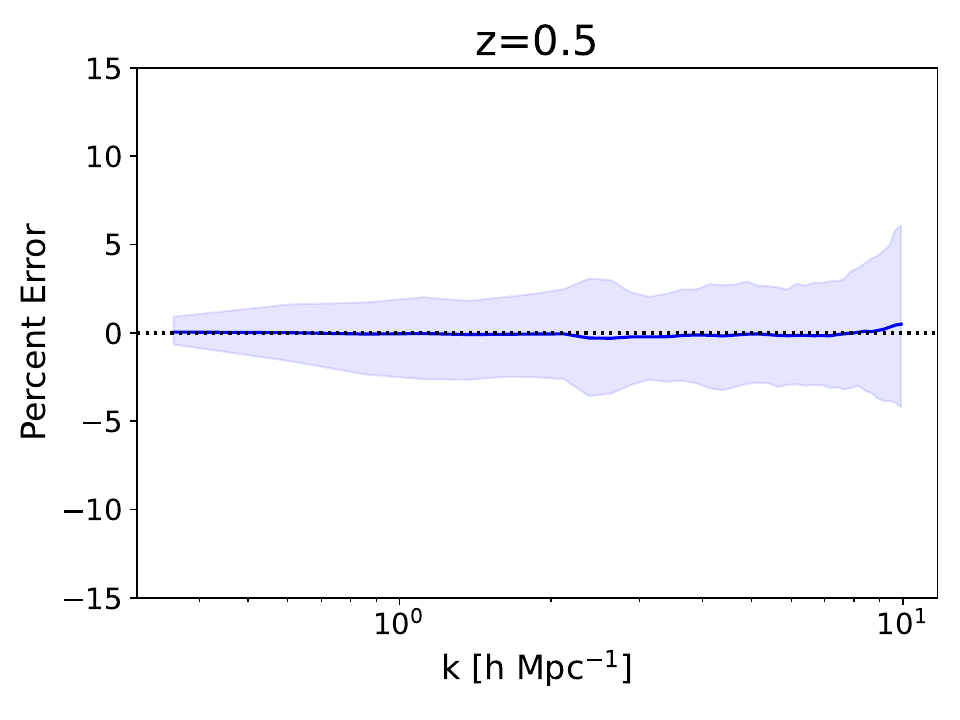}
\end{subfigure}

\begin{subfigure}{\columnwidth}
  \includegraphics[width=1.1\linewidth]{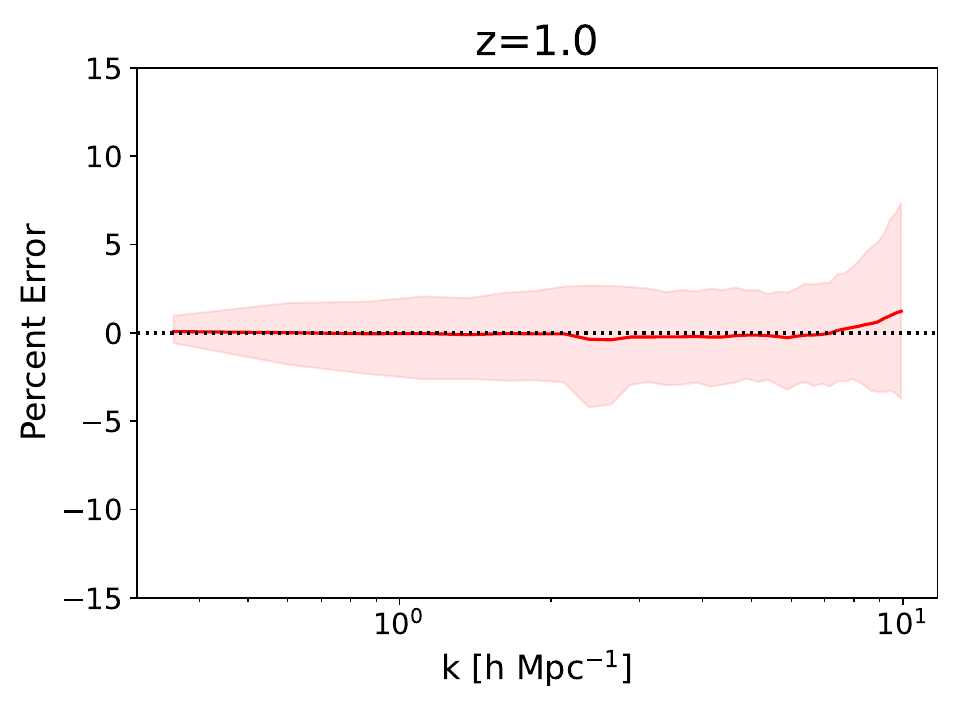}
\end{subfigure}

\begin{subfigure}{\columnwidth}
  \includegraphics[width=1.1\linewidth]{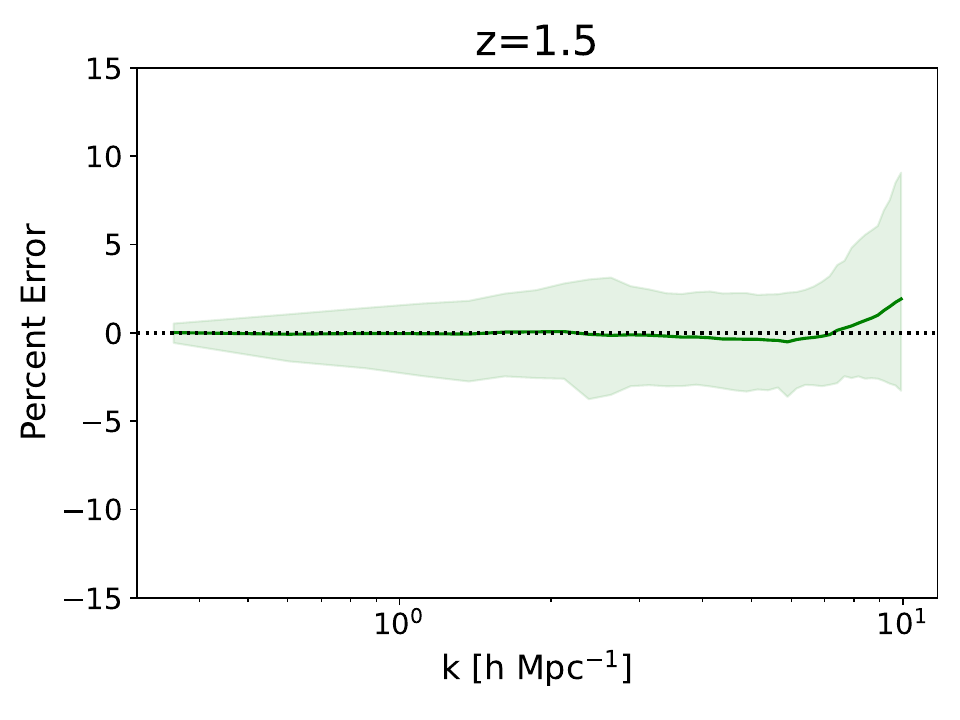}
\end{subfigure}

\columnbreak  

\begin{subfigure}{\columnwidth}
  \includegraphics[width=1.1\linewidth]{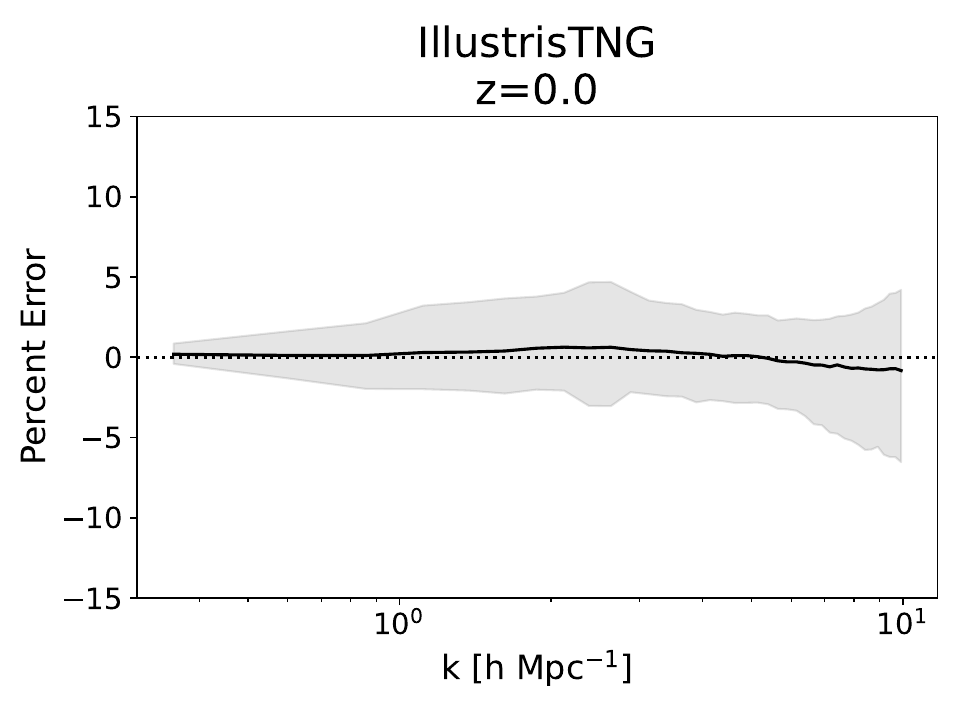}
\end{subfigure}

\begin{subfigure}{\columnwidth}
  \includegraphics[width=1.1\linewidth]{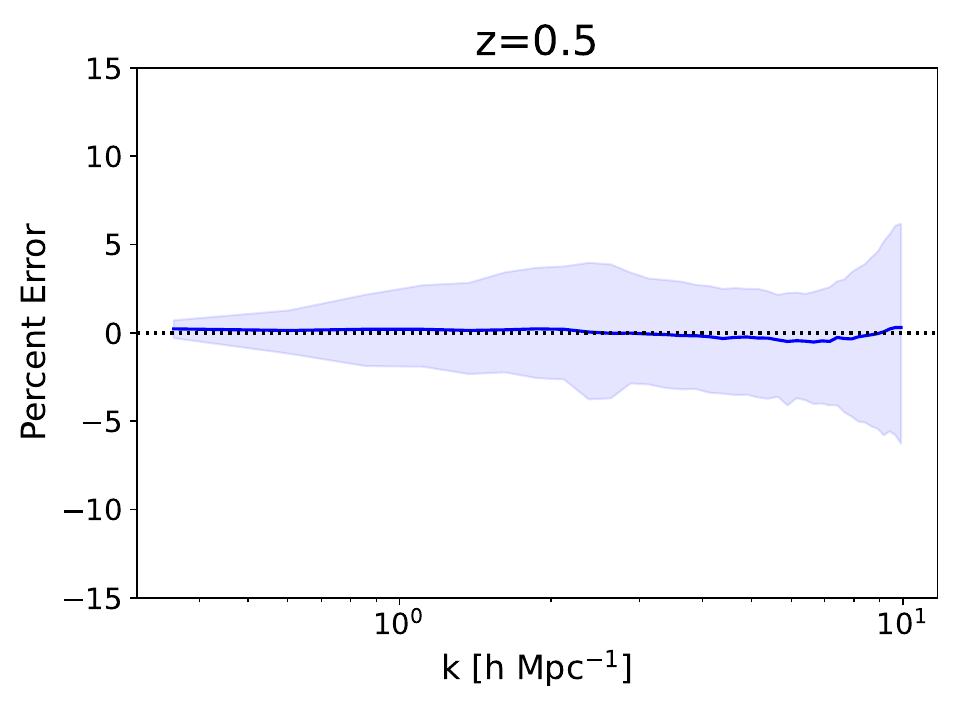}
\end{subfigure}

\begin{subfigure}{\columnwidth}
  \includegraphics[width=1.1\linewidth]{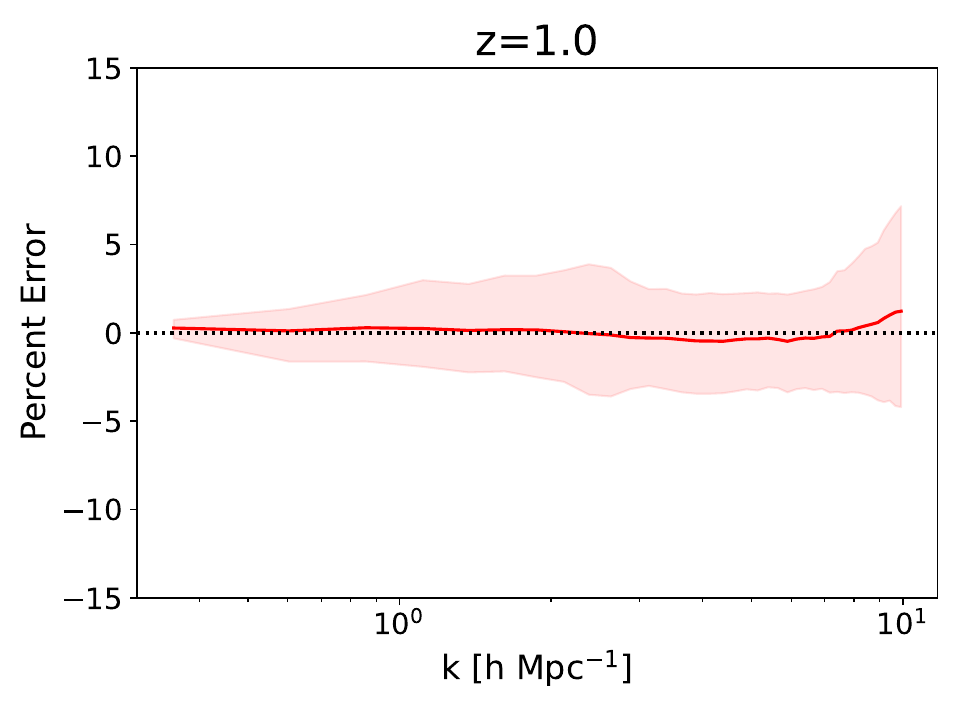}
\end{subfigure}

\begin{subfigure}{\columnwidth}
  \includegraphics[width=1.1\linewidth]{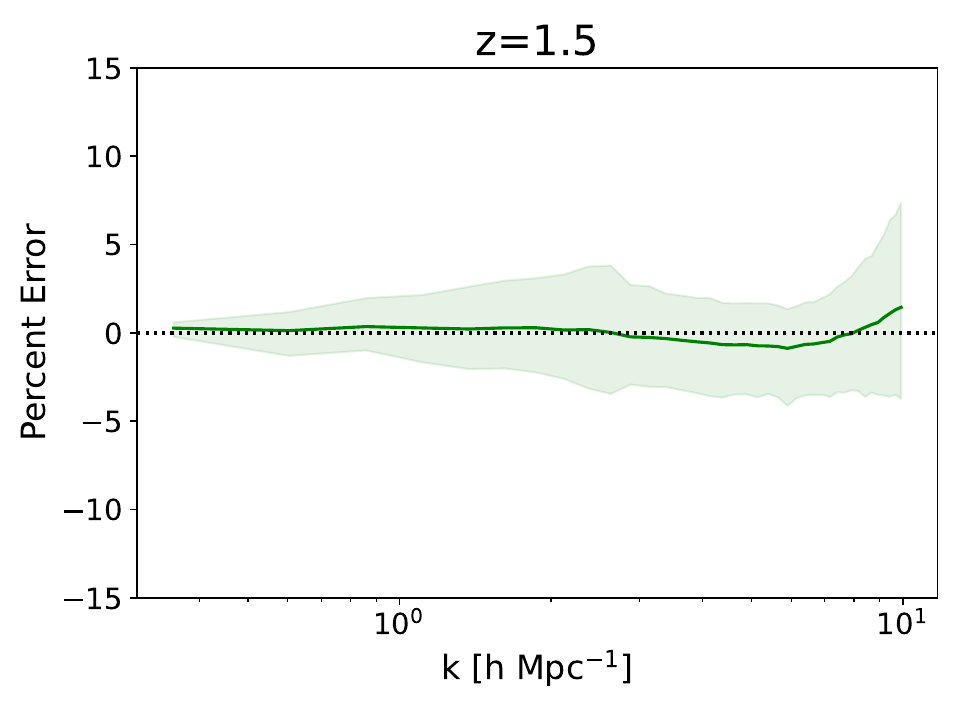}
\end{subfigure}

\columnbreak  

\begin{subfigure}{\columnwidth}
  \includegraphics[width=1.1\linewidth]{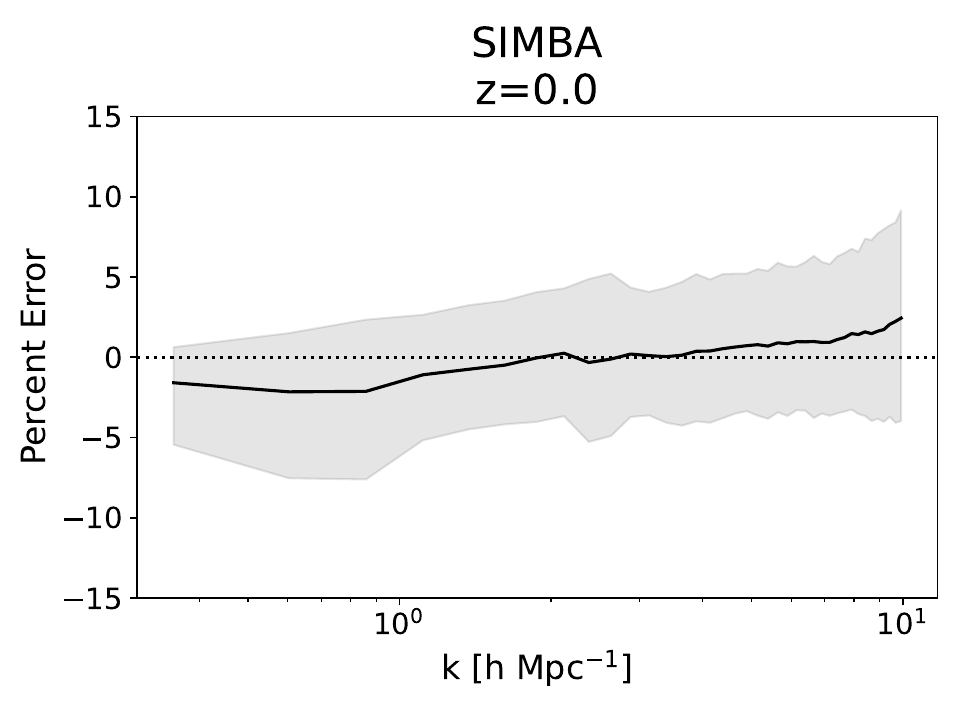}
\end{subfigure}

\begin{subfigure}{\columnwidth}
  \includegraphics[width=1.1\linewidth]{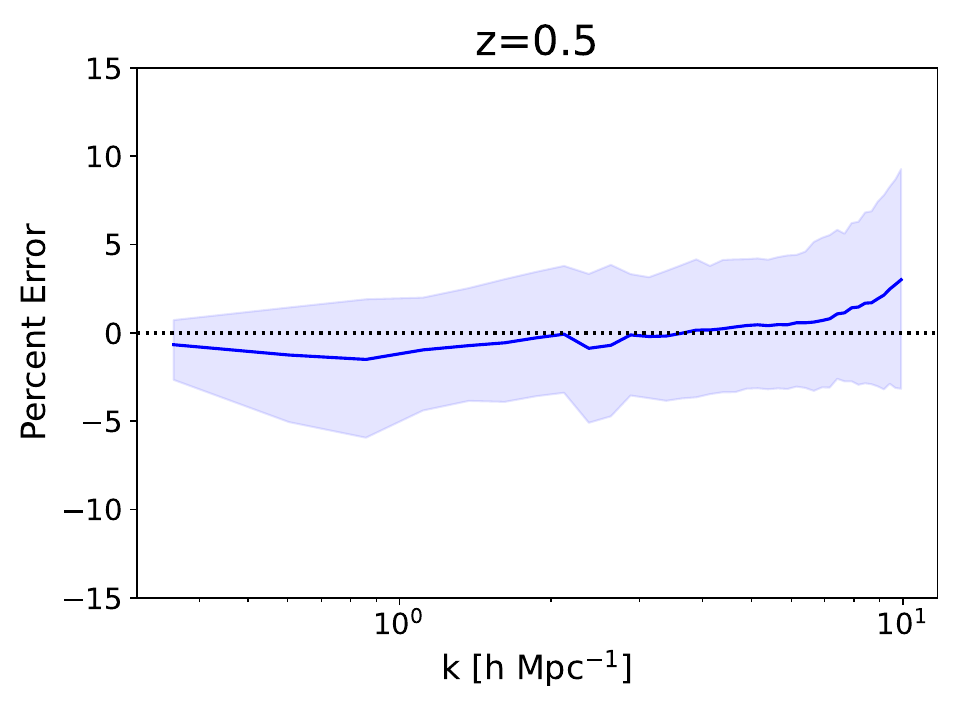}
\end{subfigure}

\begin{subfigure}{\columnwidth}
  \includegraphics[width=1.1\linewidth]{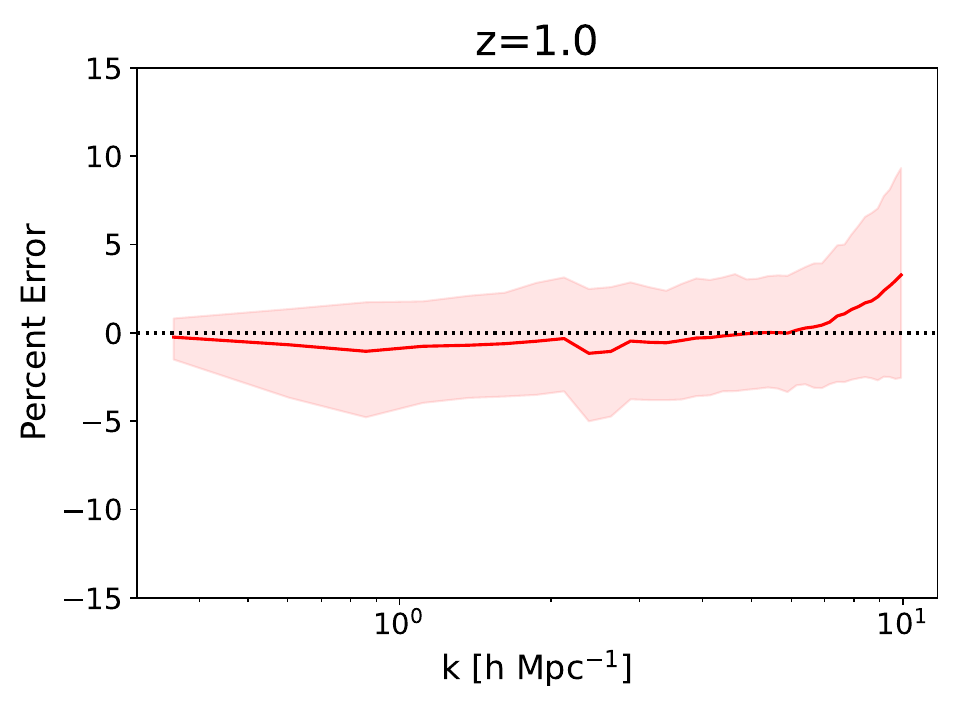}
\end{subfigure}

\begin{subfigure}{\columnwidth}
  \includegraphics[width=1.1\linewidth]{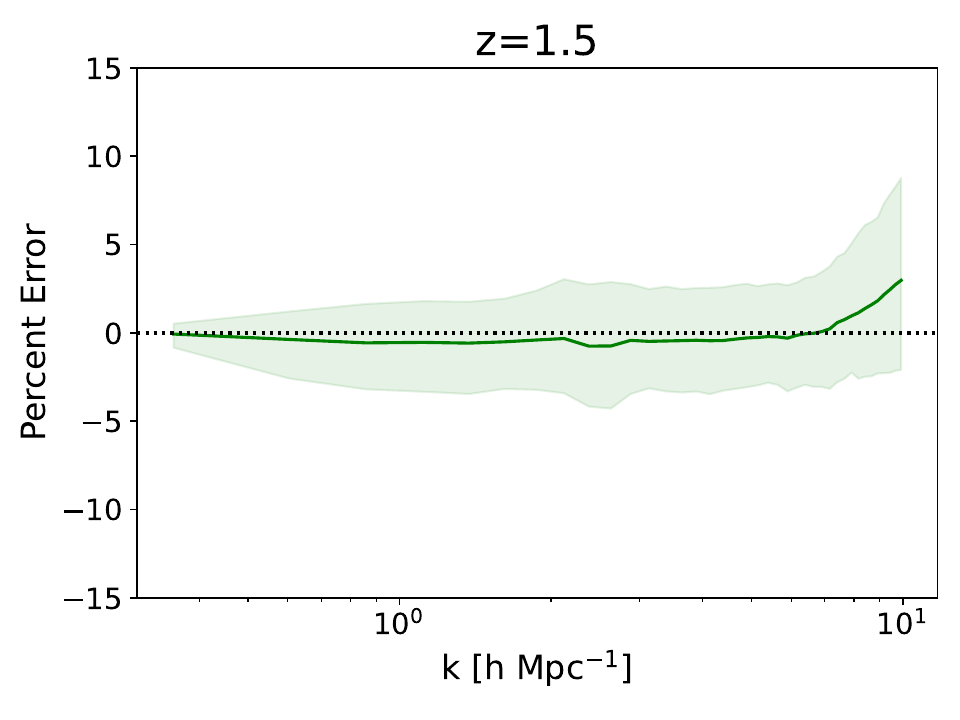}
\end{subfigure}
\end{multicols}
\caption{Emulator performance on all CAMELS test simulations. The percent error, calculated as $(\text{True Value} - \text{Predicted Value})\times 100\%$, of emulator predictions is shown as a function of wavenumber. The solid line represents the average percent error across simulations, while the shaded region denotes the 90th percentile range. Notably, the emulator demonstrates exceptional predictive accuracy, with predictions of true baryonic effects consistently achieving accuracy at the percent level. These accuracy results show that the emulator predictions are robust to changes in redshift and hydrodynamic simulation.}
\label{fig:CAMELSTestResults}
\end{figure*}

\begin{figure*}
\centering
\begin{multicols}{2} 
\begin{subfigure}{\columnwidth}
  \includegraphics[width=\linewidth]{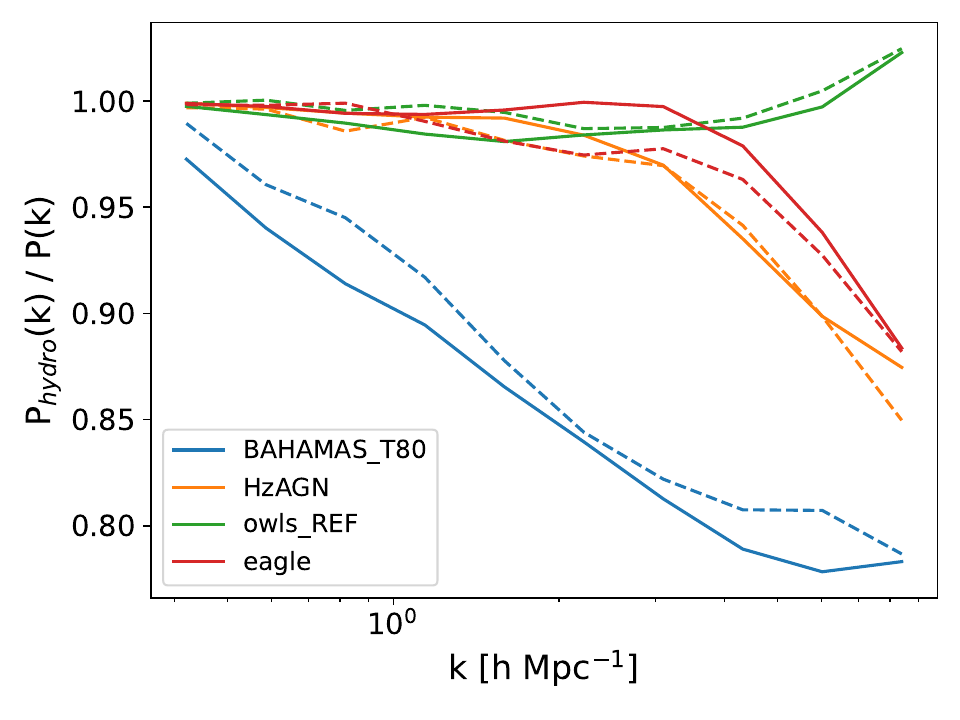}
\end{subfigure}

\begin{subfigure}{\columnwidth}
  \includegraphics[width=\linewidth]{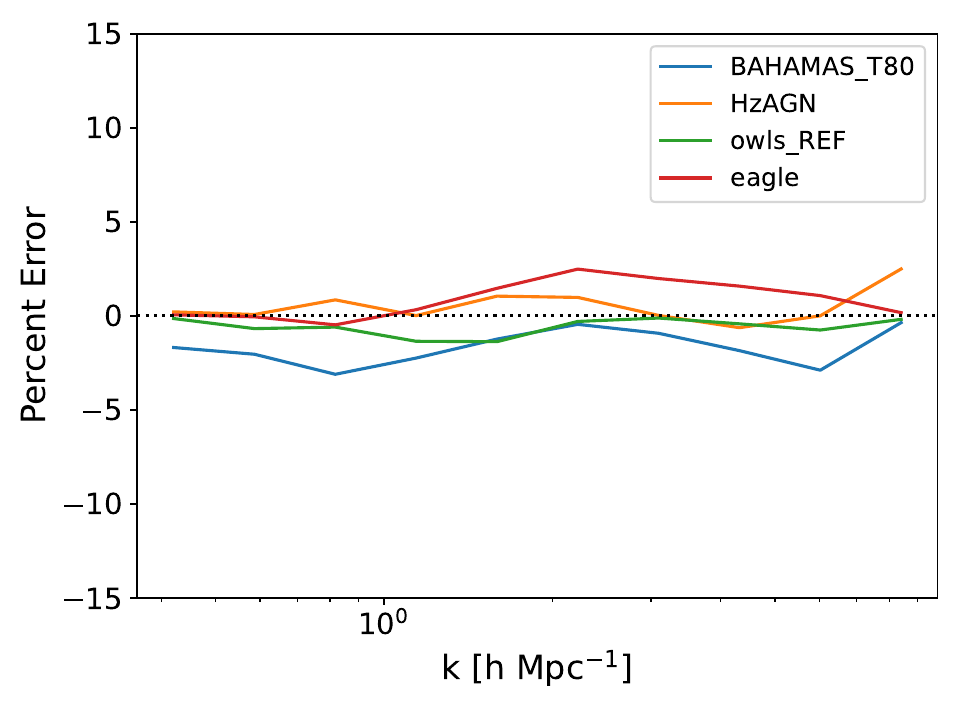}
\end{subfigure}
\end{multicols}
\caption{The emulator predictions alongside the reference hydrodynamic simulations for scenarios outside the CAMELS dataset.
The \textbf{left panel} shows the matter power spectrum ratios. Each solid line represents the ground truth from the respective hydrodynamic simulation, while the corresponding dashed line showcases the emulator's predictions. Notably, the emulator closely matches the underlying behavior of the baryonic effects in these distinct simulations.
The \textbf{right panel} shows the prediction errors for each simulation. The emulator consistently achieves accuracy at the percent level, accurately capturing the baryonic effects in these diverse scenarios. The predictions demonstrate robustness even amidst variations in hydrodynamic simulations, highlighting the emulator's capability to adapt and predict the impact of baryonic physics with high accuracy across a broad spectrum of simulations outside the CAMELS dataset.}
\label{fig:OtherHydroTestResults}
\end{figure*}

\begin{itemize}
\item \textbf{CAMELS simulations.} Figure \ref{fig:CAMELSTestResults} shows the error achieved by our emulator for simulations of three different suites of CAMELS (IllustrisTNG, Astrid, and SIMBA) at four different redshifts. The solid lines represent the average percent error across simulations, while the shaded regions denote the 90th percentile range. These results correspond to all the baryonic effects illustrated in figure \ref{fig:CAMELSRatios}.
Firstly, we can see that the emulator achieves a high accuracy down to $k \sim 10$ h/Mpc with deviations remaining typically less than 5\%.  We emphasize that our emulator is robust to changes in redshifts and baryonic effects across CAMELS.

The performance of the emulator is similar across redshifts for the Astrid and IllustrisTNG simulations at all scales, with higher accuracy at large scales and somewhat lower precision at smaller scales. However, at $z=0.0$  and $0.5$, the prediction error for SIMBA can be as high $\sim 5\%$ on large scales. This is likely due to the aggressive AGN feedback in SIMBA, which produces the most prominent suppression of the matter power on large scales \citep{gebhardt2023cosmological} as seen even in the comparison plots in Figure \ref{fig:CAMELSRatios}. Nonetheless, the prediction error is still within $\sim 5\%$ and is comparable to the other two suites at $z=1.0, 1.5$.

\item \textbf{Non-CAMELS simulations.} While the above test shows the robustness of our emulator to changes in cosmology, astrophysics, and subgrid physics, we note that all CAMELS simulations share the same volume and resolution. In order to quantify how well our emulator behaves to changes in volume, resolution, and other subgrid physics models, we quantify how well it is able to reproduce the results of the  BAHAMAS, Horizon AGN, Owls, and Eagle simulations. We show the results in Figure \ref{fig:OtherHydroTestResults}. 

The left panel shows the correction to the matter power spectrum in these simulations; solid lines represent the simulation results, while the corresponding dashed line depicts our emulator's predictions. The emulator closely mirrors the inherent behavior of baryonic effects across these varied simulations.  In the right panel, the prediction errors for each simulation are displayed. Consistently, the emulator maintains accuracy at the percent level up to $k \sim 10$ h/Mpc, adeptly capturing the intricacies of baryonic effects across diverse scenarios. All the above tests clearly illustrate the versatility and robustness of our emulator, which is capable of reproducing the ratio $P_{\rm hydro}(k)/P_{\rm nbody}(k)$ for thousands of simulations with different cosmologies, astrophysics, subgrid physics, volumes, resolutions, and redshifts.

\end{itemize}

\subsection{Comparison against other emulators}
In recent years, different groups have created emulators to model baryonic effects for 2-point statistics. Given the findings of this work, we can also use those emulators to create field-level baryonic effects corrections. In this subsection, we conduct comparative evaluations against widely used emulators such as BACCO, HMcode, and BCemu, both within and beyond the CAMELS simulations. Through the following comparisons, we show that, overall, our emulator offers greater flexibility and robustness in modeling baryonic effects compared to the other emulators.

\begin{figure*} 
\centering
\begin{multicols}{3} 
\begin{subfigure}{\columnwidth}
  \includegraphics[width=1.1\linewidth]{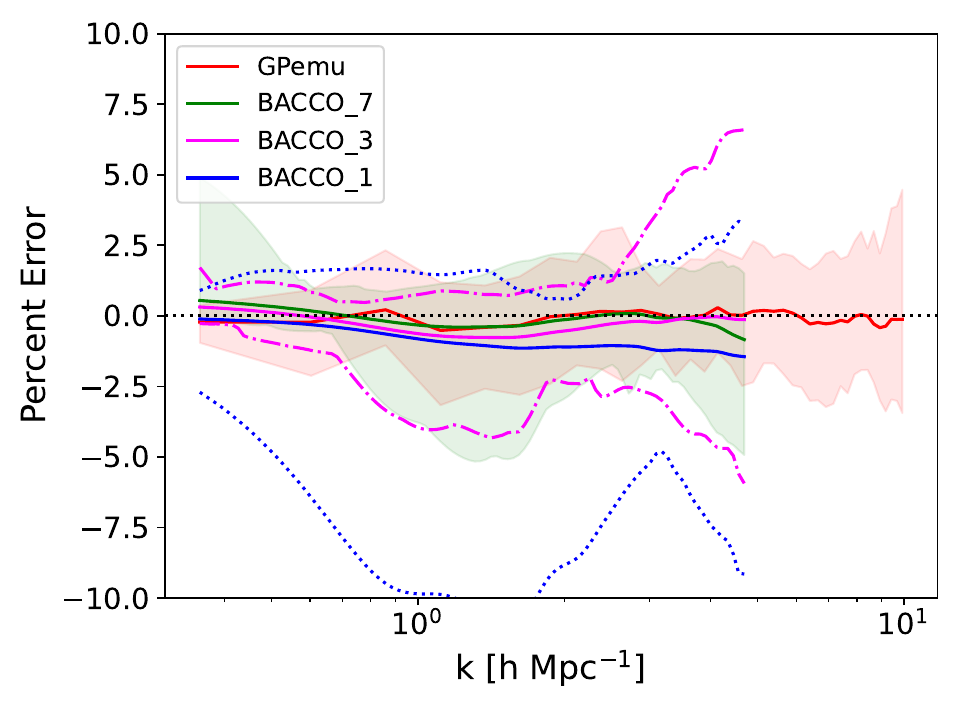}
\end{subfigure}

\begin{subfigure}{\columnwidth}
  \includegraphics[width=1.1\linewidth]{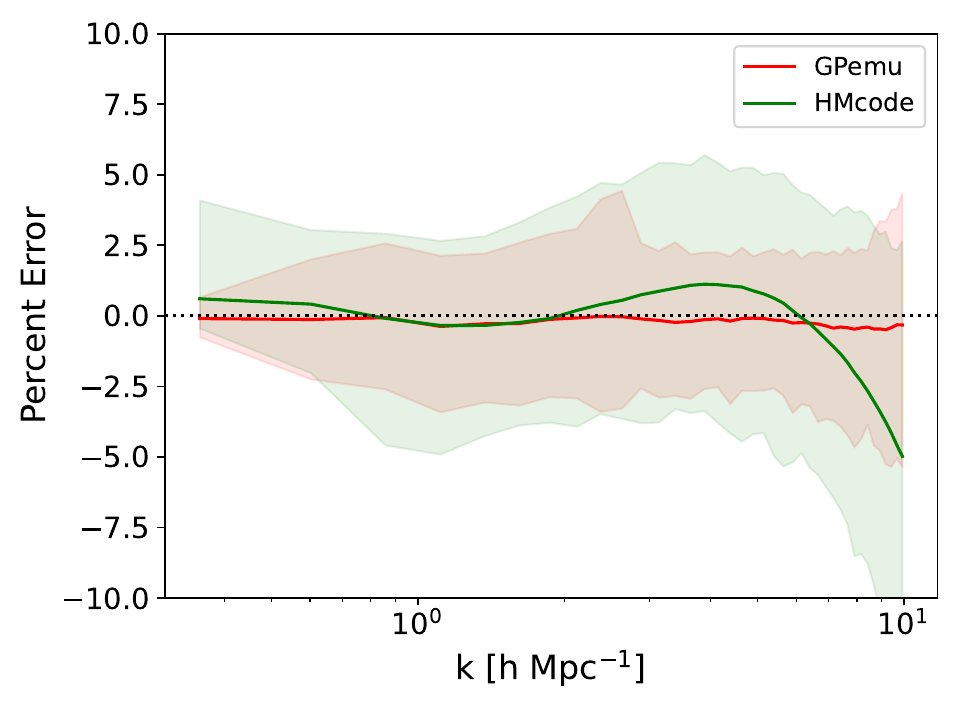}
\end{subfigure}

\begin{subfigure}{\columnwidth}
  \includegraphics[width=1.1\linewidth]{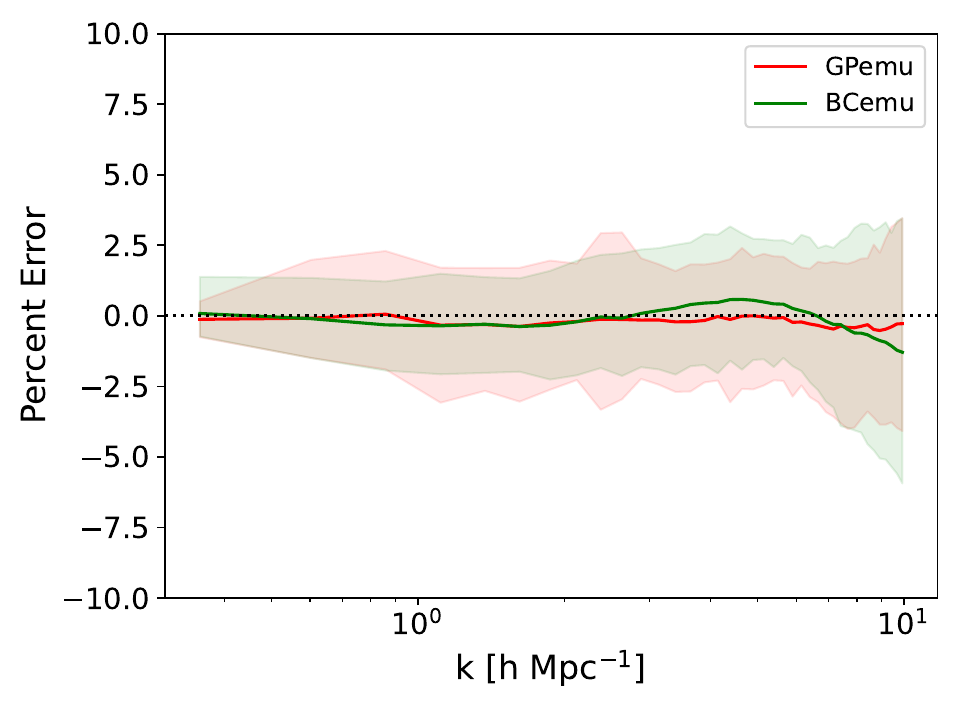}
\end{subfigure}
\end{multicols}
\caption{Comparison between our GP emulator and other baryonification emulators - BACCO (left panel), HMcode (middle panel), and BCemu (right panel) - on Astrid test data at $z=0.0$. The solid lines represent the average percent error, while the shaded regions depict the 90th percentile of errors.
\textbf{Left Panel:} Within BACCO's acceptable cosmological range, we use  39 test simulations for emulation, covering up to its maximum wavenumber $k \sim 5$. The dash-dotted and dotted lines illustrate the 90th percentile outputs for BACCO's 3 and 1 parameter emulators, respectively. Our GP emulator demonstrates the capability to investigate smaller scales, exhibiting accuracy comparable to BACCO's 7-parameter model while offering increased flexibility and generality.
\textbf{Middle Panel:} Using all 200 test simulations, our GP emulator demonstrates comparable performance to HMcode on larger scales and exhibits higher accuracy on smaller scales.
\textbf{Right Panel:} Using the 148 test simulations that fall within BCemu's acceptable cosmological range, both emulators exhibit comparable performance, but our GP emulator showcases greater flexibility and generality in its predictions.
}
\label{fig:CAMELS_emu_comparison}
\end{figure*}

\begin{figure*} 
\centering
\begin{multicols}{4} 

\begin{subfigure}{\columnwidth}
  \includegraphics[width=1.2\linewidth]{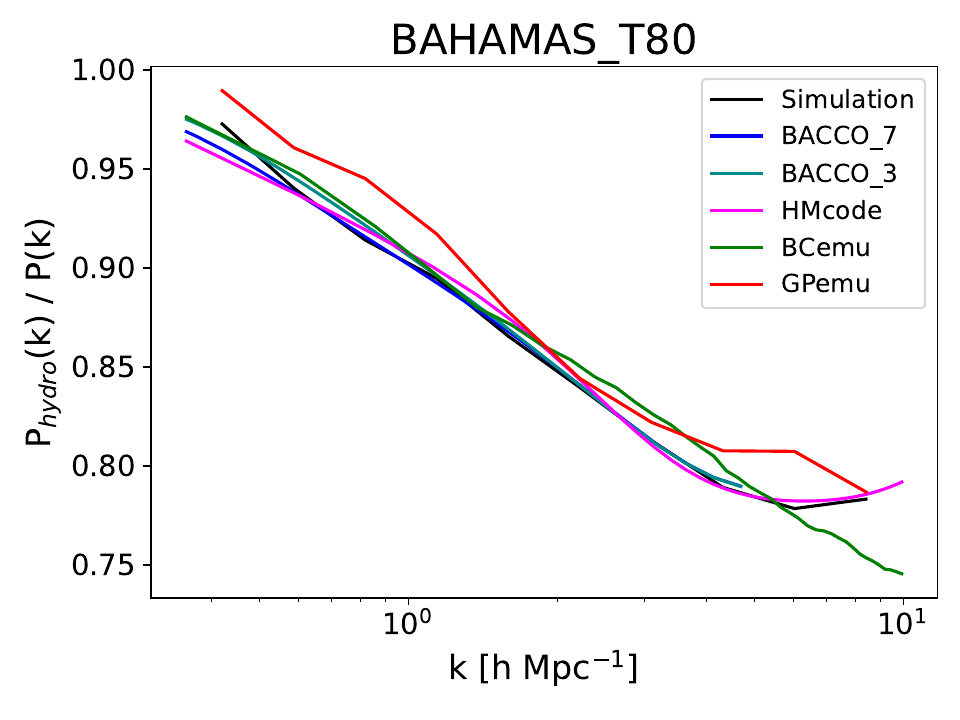}
\end{subfigure}

\begin{subfigure}{\columnwidth}
  \includegraphics[width=1.2\linewidth]{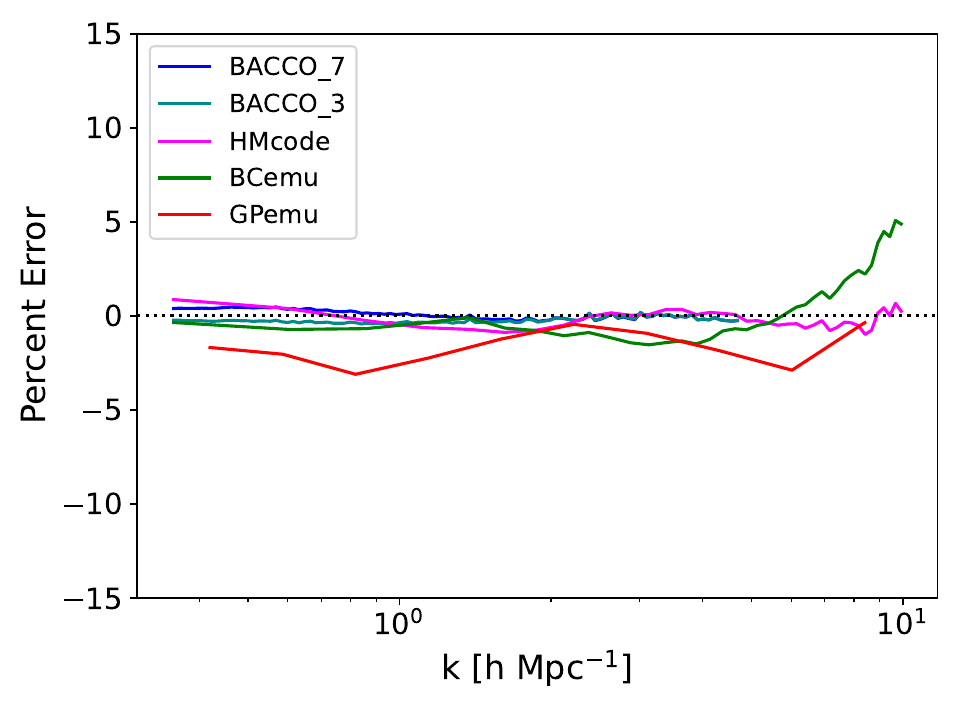}
\end{subfigure}

\columnbreak  

\begin{subfigure}{\columnwidth}
  \includegraphics[width=1.2\linewidth]{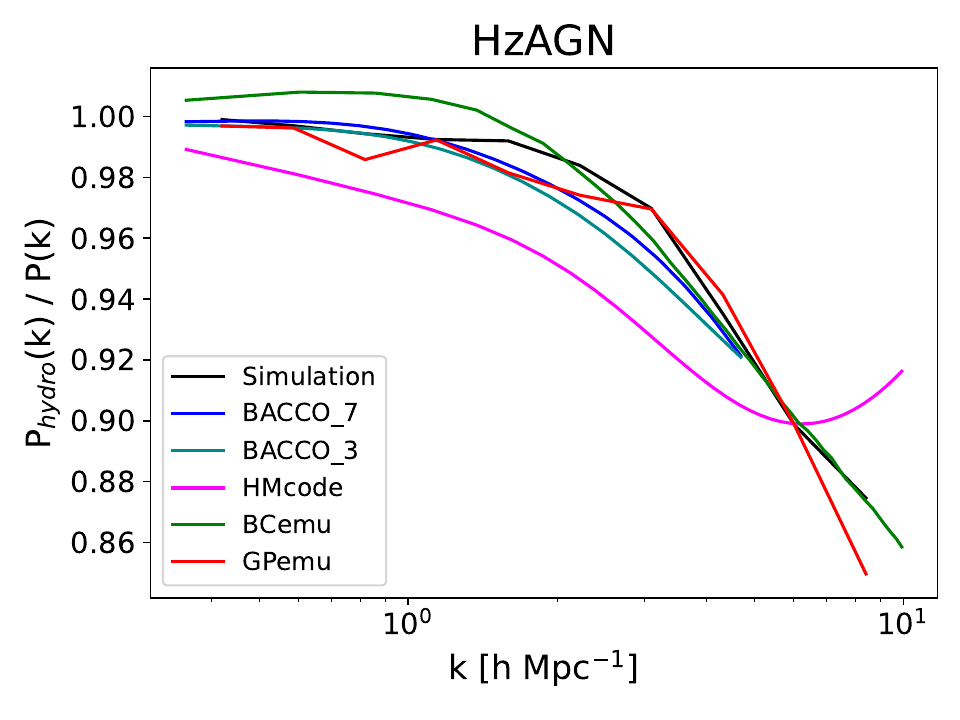}
\end{subfigure}

\begin{subfigure}{\columnwidth}
  \includegraphics[width=1.2\linewidth]{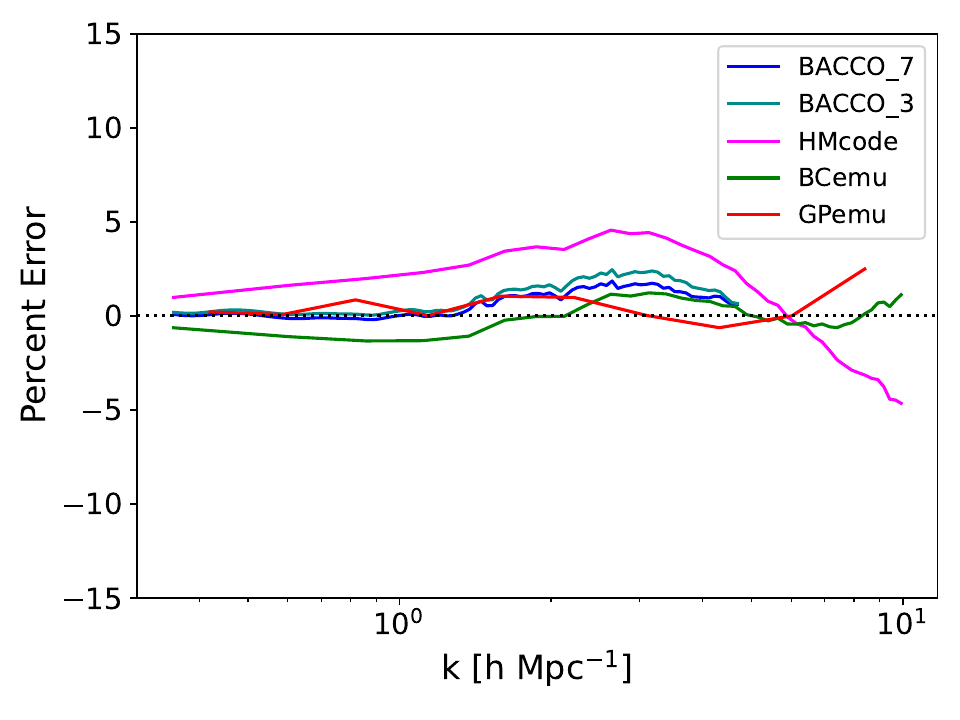}
\end{subfigure}

\columnbreak  

\begin{subfigure}{\columnwidth}
  \includegraphics[width=1.2\linewidth]{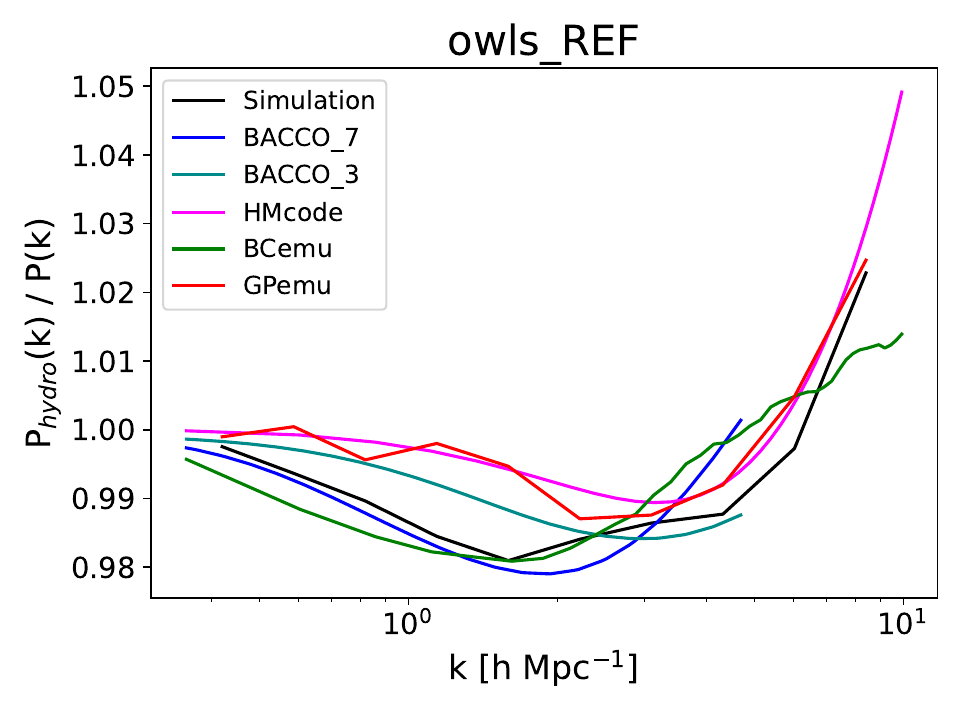}
\end{subfigure}

\begin{subfigure}{\columnwidth}
  \includegraphics[width=1.2\linewidth]{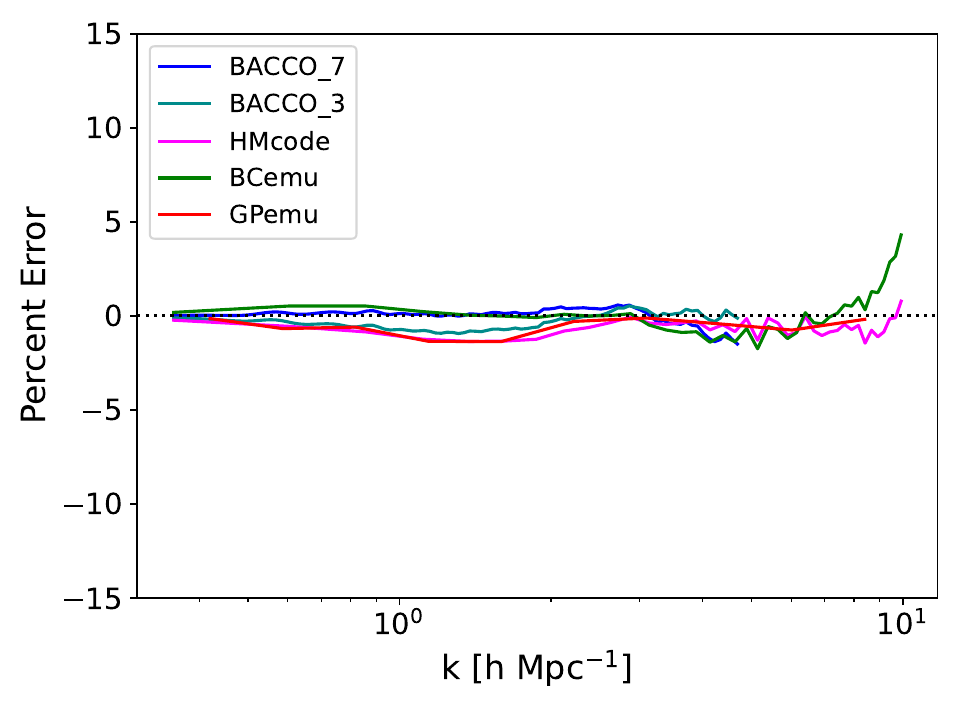}
\end{subfigure}

\columnbreak  

\begin{subfigure}{\columnwidth}
  \includegraphics[width=1.2\linewidth]{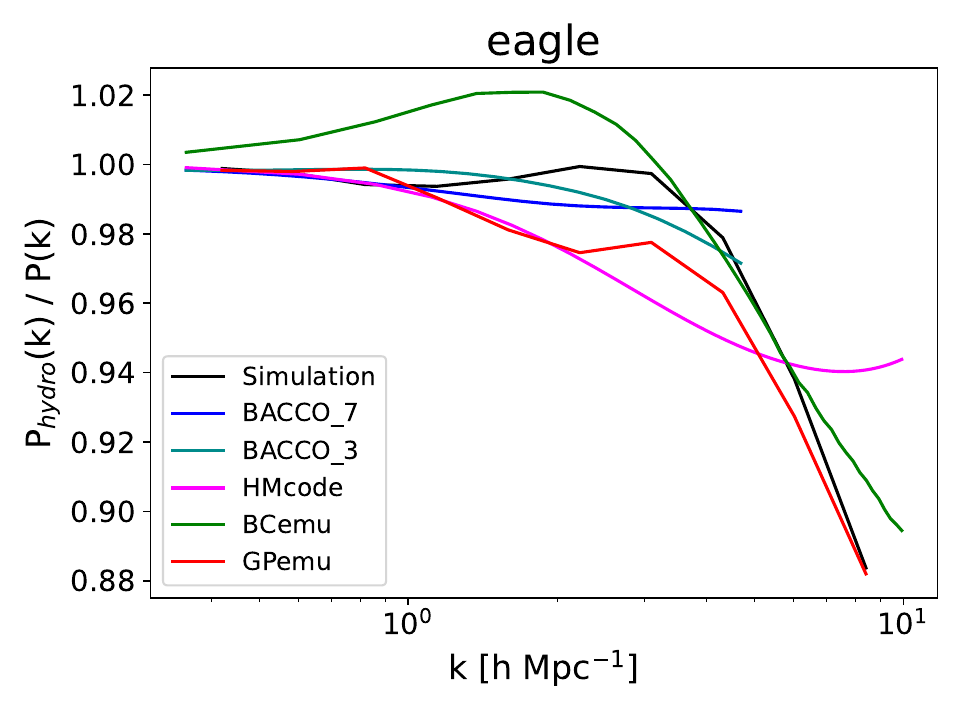}
\end{subfigure}

\begin{subfigure}{\columnwidth}
  \includegraphics[width=1.2\linewidth]{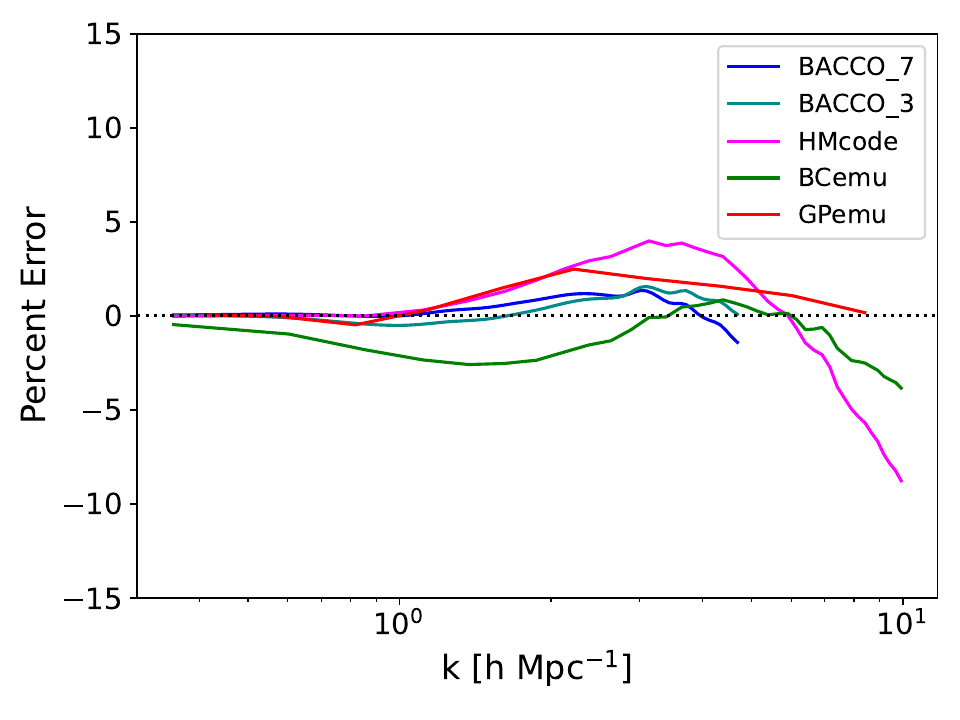}
\end{subfigure}
\end{multicols}
\caption{Comparison of our emulator against BACCO's 7 and 3 parameter versions, HMcode, and BCemu using hydrodynamic simulations outside of CAMELS. The simulations encompass varying degrees of baryonic feedback. For BACCO, emulation is performed up to BACCO's maximum wavenumber of $k \sim 5$. Despite differences in the length scales of emulation, all emulators achieve percent-level accuracy in predicting baryonic effects, showcasing comparable performance within their respective length scales.}
\label{fig:BACCO+HMcomparisonOtherHydro}
\end{figure*}

\begin{itemize}

\item \textbf{BACCO:} BACCO is a neural network-based emulator that accounts for baryonic effects in the non-linear matter power spectrum \citep{BACCOpaper}. BACCO encompasses a parameter set comprising 8 cosmological parameters, consisting of the standard 5 $\Lambda$CDM parameters combined with massive neutrinos and dynamical dark energy. Additionally, it includes 7 free baryonic parameters derived from physical principles, describing factors such as the gas fraction retained in halos, the intensity of AGN feedback, the characteristic galaxy mass, and the relationship between gas fractions and halo mass. In addition to the 7 free parameter model, BACCO also has 3 and 1 parameter models. When not included in the model, the baryonic parameters are fixed at their fiducial values. BACCO achieves an overall precision of $\sim$ 1-5\% across its models and its targeted scales ($0.01 < k < 5$ h/Mpc) and redshifts ($0 < z < 1.5$), encompassing various cosmological hydrodynamic simulations. However, BACCO's capacity to confidently predict the baryon-corrected power spectrum is limited to a maximum wavenumber of k = 4.7 h/Mpc, notably smaller than our emulator's range. Furthermore, its range of validity is narrower than GPemu:  $\sigma_8 \in [0.73, 0.9]$ and $\Omega_m \in [0.23, 0.4]$. As a result, only 39 out of the 200 Astrid $z=0.0$ test simulations are within BACCO's specified cosmology range. 

The left panel of Figure \ref{fig:CAMELS_emu_comparison} shows the comparison between BACCO's predictions (including the 7, 3, and 1 parameter models) and our emulator's predictions on these limited 39 simulations. The solid lines represent the average percent error, while the shaded regions depict the 90th percentile of errors. The dash-dotted and dotted lines illustrate the 90th percentile outputs for BACCO’s 3 and 1 parameter emulators, respectively. The comparison results for the SIMBA and IllustrisTNG suites are similar. In Figure \ref{fig:BACCO+HMcomparisonOtherHydro} we compare GPemu against BACCO for the non-CAMELS simulations. 

Overall, we find that GPemu exhibits an accuracy similar to that of BACCO, but its range of validity, both in terms of scales and parameter-space, is wider.


\item \textbf{HMcode:} The HMcode \citep{mead2021hmcode, HMcode} is a simple halo model designed to simulate the influence of baryonic feedback on the power spectrum. It incorporates a six-parameter physical framework that includes gas expulsion by AGN feedback and encapsulates star formation. The feedback model was fitted to simulation data, taken from the library of \citet{vanDaalen_2020}. 

In our evaluation, similar to the comparison conducted against BACCO, we conducted a side-by-side analysis of HMcode's predictions alongside our emulator's outcomes using Astrid test data. The results of this comparison are illustrated in the middle panel of Figure \ref{fig:CAMELS_emu_comparison}, with solid lines representing the average percent error and shaded regions depicting the 90th percentile of the errors. We can see that our emulator demonstrates comparable performance
to HMcode on larger scales while exhibiting higher accuracy on smaller scales where baryonic effects are stronger. A similar conclusion can be reached by comparing HMCode against GPemu for non-CAMELS simulations as shown in Figure \ref{fig:BACCO+HMcomparisonOtherHydro}. 

\item \textbf{BCemu:} The BCemu emulator \citep{giri2023bcemu} focuses on modeling the baryonic suppression of the matter power spectrum. It is based on a slightly modified version of the baryonification model \citep{Schneider_2019} and features seven physically-meaningful free-parameters related to gas profiles and stellar abundances within halos. BCemu demonstrated its capability to replicate the power spectra of hydrodynamical simulations with sub-percent precision. Moreover, it established a correlation between the baryonic suppression of the power spectrum and the gas and stellar fractions within halos. However, similar to BACCO, BCemu is constrained by its limited acceptance range for cosmological parameters ($\Omega_m \in [0.196, 0.49]$), encompassing only 148 out of the 200 Astrid test simulations. 

The right panel of Figure \ref{fig:CAMELS_emu_comparison} compares BCemu's predictions with those of our emulator within this subset, with solid lines representing the average percent error and shaded regions depicting the 90th percentile of the errors. From Figure \ref{fig:BACCO+HMcomparisonOtherHydro} we can see that GPemu performs similarly to BCemu when used on non-CAMELS simulations. While both emulators display comparable performance at all scales, our GP emulator shows greater flexibility and generality in its predictions of baryonic effects across a wider range of hydrodynamic simulations.

\end{itemize}

\subsection{Field-level emulation}

From Figure \ref{fig:CAMELS_CC} we found that baryonic effects do not significantly affect the phases of Fourier modes down to $k\sim10~h{\rm Mpc}^{-1}$. Thus, baryonic effects at the field level can be accounted for by correcting the amplitude of Fourier modes from N-body simulations. Now that we have an emulator for the ratio, $S(k)=P_{\rm hydro}(k)/P_{\rm nbody}(k)$, we can investigate how well our model performs at the field level.
The resulting field-level transformations exhibit effective improvements, evident across multiple simulation suites at different redshifts.

In more detail, the procedure we employ to model baryonic effects at the field level is as follows. First, we take a given hydrodynamic simulation and its N-body counterpart. We then compute the power spectrum of each of them to compute the baryonic suppression: $S(k)=P_{\rm hydro}(k)/P_{\rm nbody}(k)$. Next, we fit the four free parameters of GPemu to get the best match to $S(k)$. Then, from the N-body simulation, we compute the matter density field $\delta_{\rm nbody}(\mathbf{x})$ and its Fourier transform: $\delta_{\rm nbody}(\mathbf{k})=A_{\mathbf{k}}e^{i\theta_{\mathbf{k}}}$. Finally, we obtain the baryon-corrected field by Fourier transforming back $\delta_{\rm postTF}(\mathbf{k})$, where
\begin{equation}
\delta_{\rm postTF}(\mathbf{k})=\sqrt{S_{\rm GPemu}(k)} \delta_{\rm nbody}(\mathbf{k})
\end{equation}
with $\sqrt{S_{\rm GPemu}(k)}$ being the transfer function predicted by our emulator GPemu.

\begin{figure*}  
\centering
\begin{multicols}{4} 
\begin{subfigure}{\columnwidth}
  \includegraphics[width=1.2\linewidth]{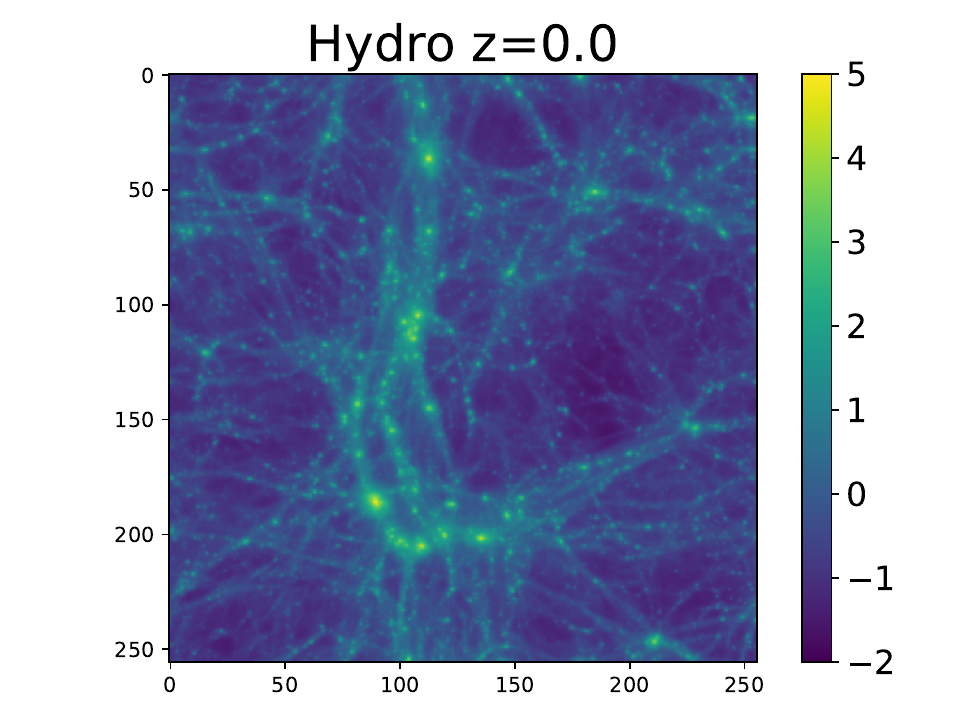}
\end{subfigure}

\begin{subfigure}{\columnwidth}
  \includegraphics[width=1.2\linewidth]{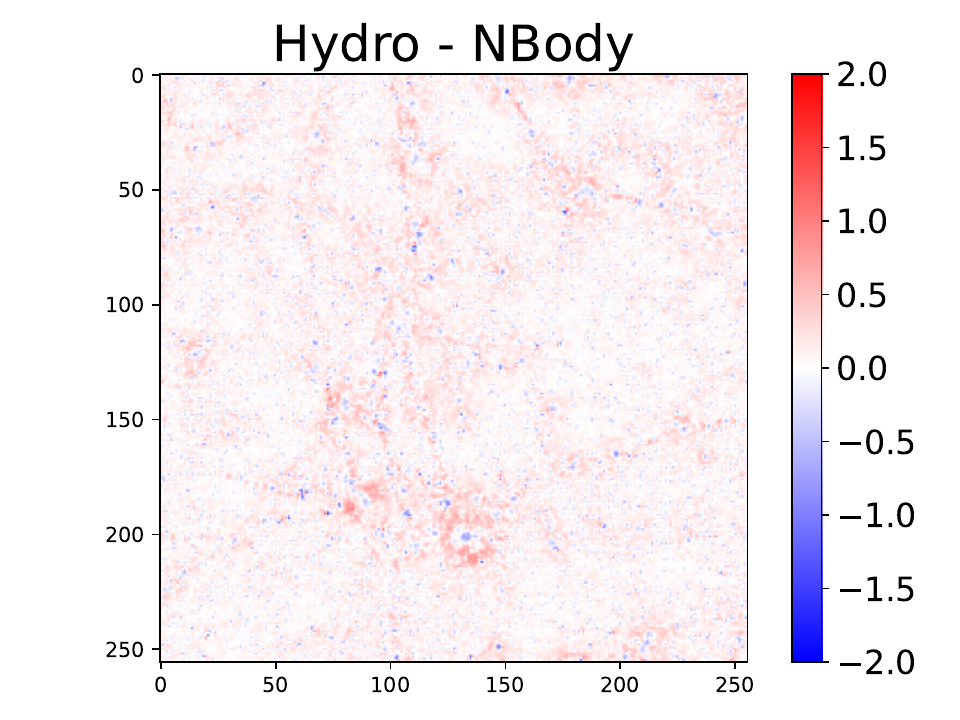}
\end{subfigure}

\begin{subfigure}{\columnwidth}
  \includegraphics[width=1.2\linewidth]{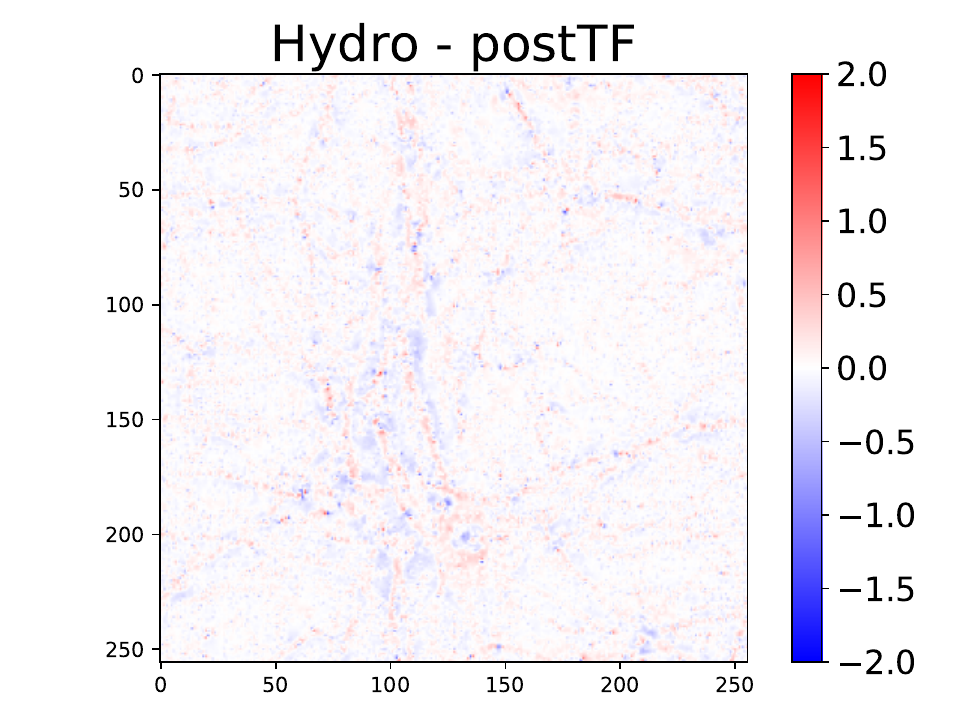}
\end{subfigure}

\begin{subfigure}{\columnwidth}
  \includegraphics[width=1.15\linewidth]{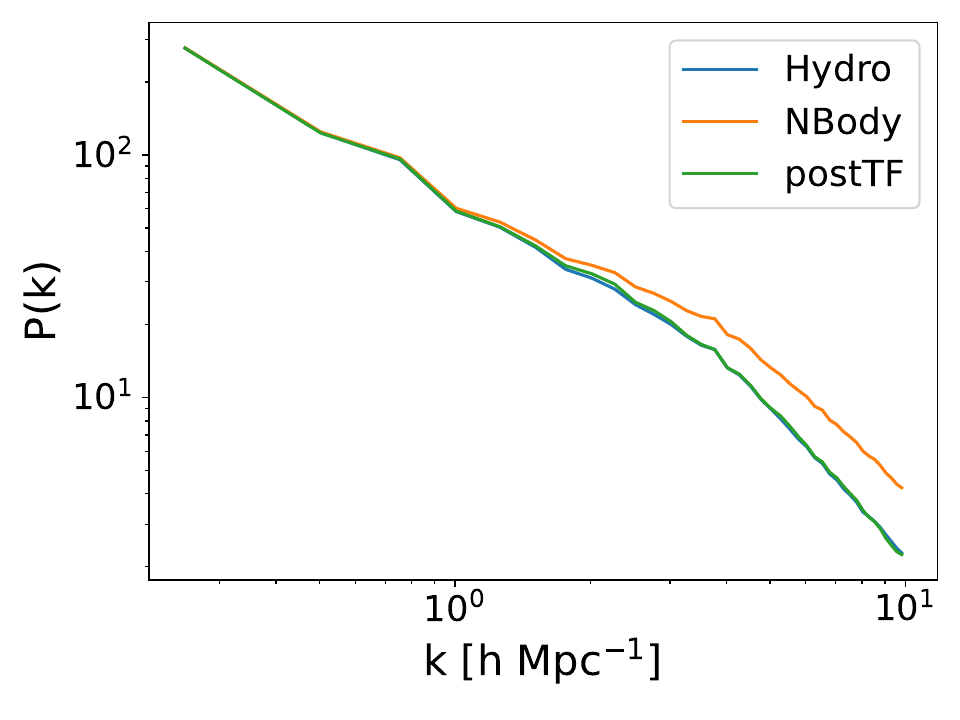}
\end{subfigure}

\columnbreak  

\begin{subfigure}{\columnwidth}
  \includegraphics[width=1.2\linewidth]{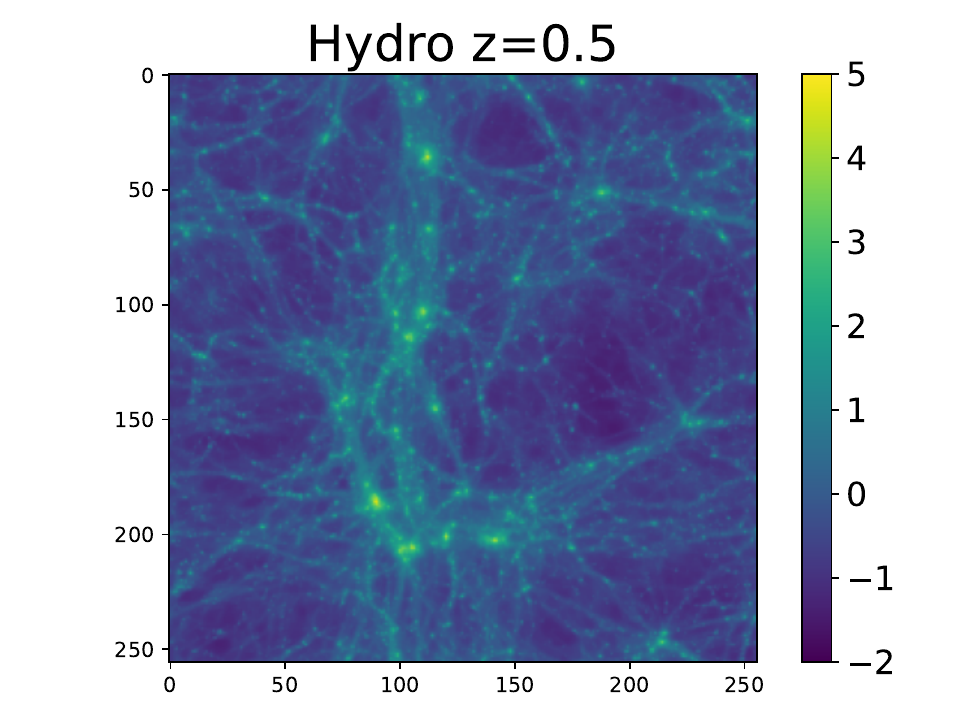}
\end{subfigure}

\begin{subfigure}{\columnwidth}
  \includegraphics[width=1.2\linewidth]{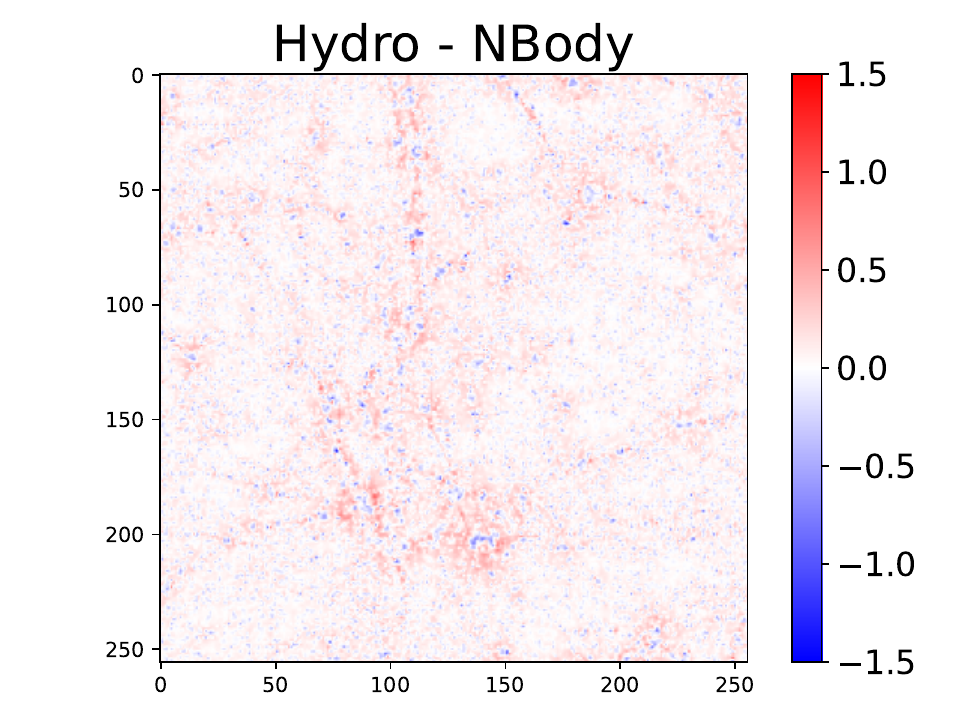}
\end{subfigure}

\begin{subfigure}{\columnwidth}
  \includegraphics[width=1.2\linewidth]{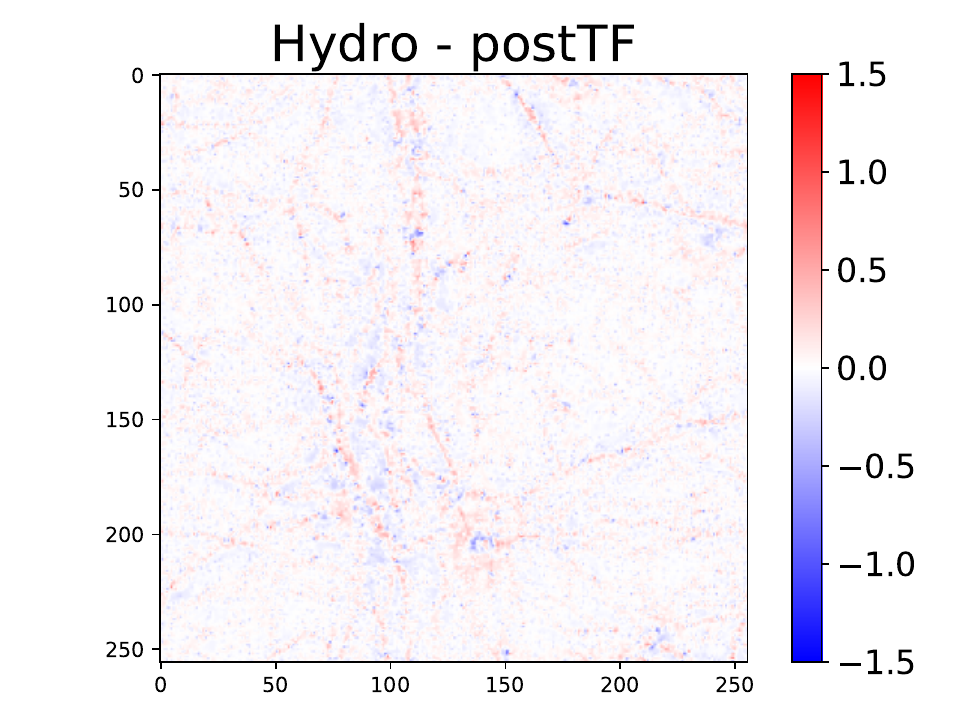}
\end{subfigure}

\begin{subfigure}{\columnwidth}
  \includegraphics[width=1.15\linewidth]{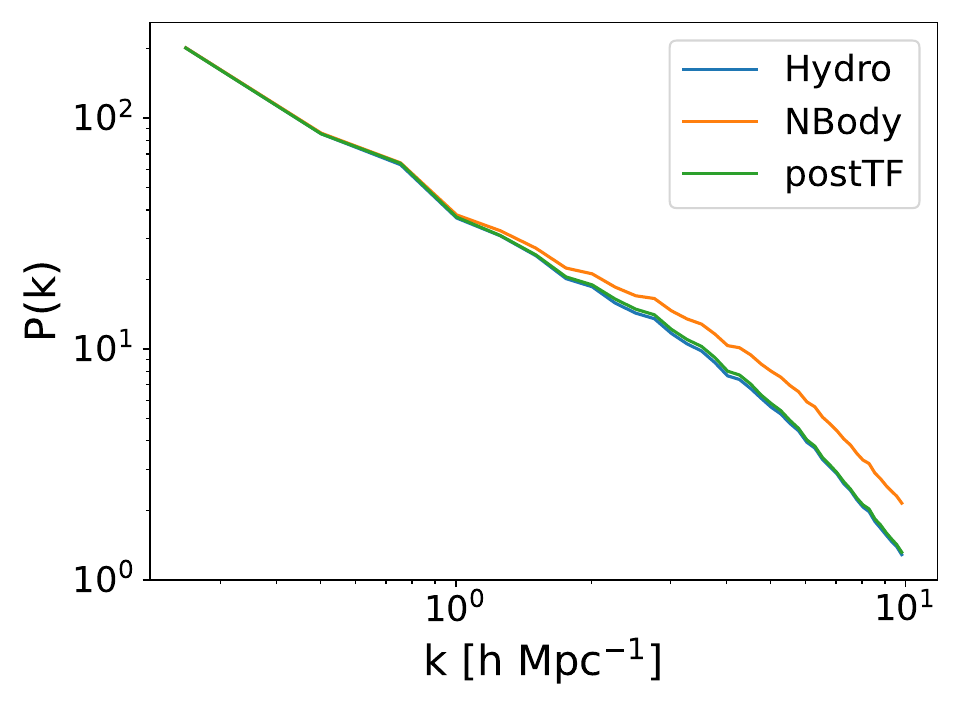}
\end{subfigure}

\columnbreak  

\begin{subfigure}{\columnwidth}
  \includegraphics[width=1.2\linewidth]{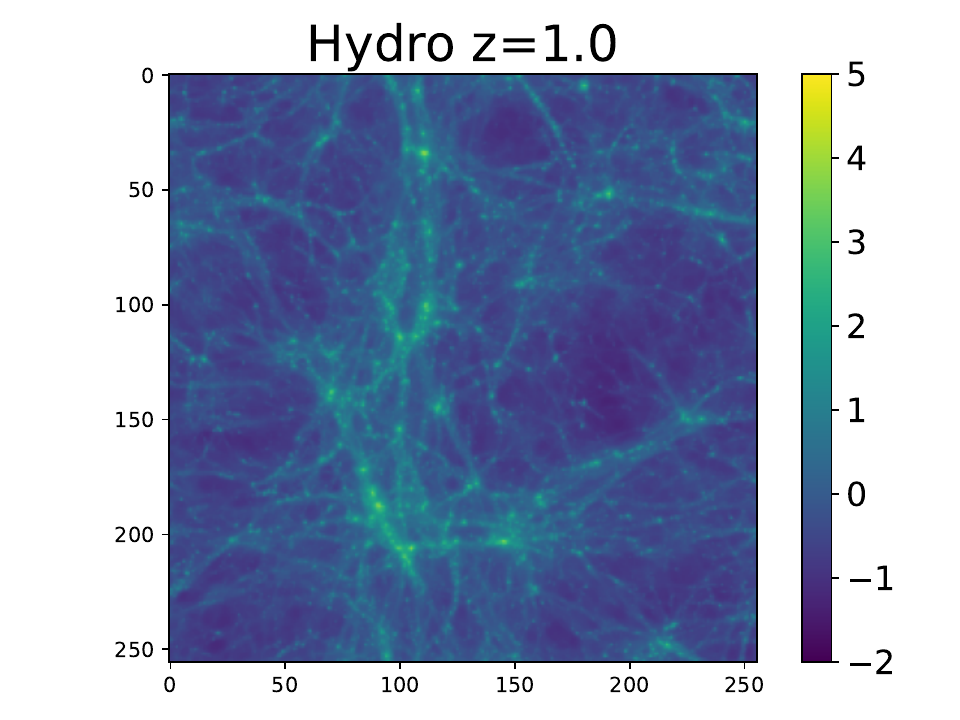}
\end{subfigure}

\begin{subfigure}{\columnwidth}
  \includegraphics[width=1.2\linewidth]{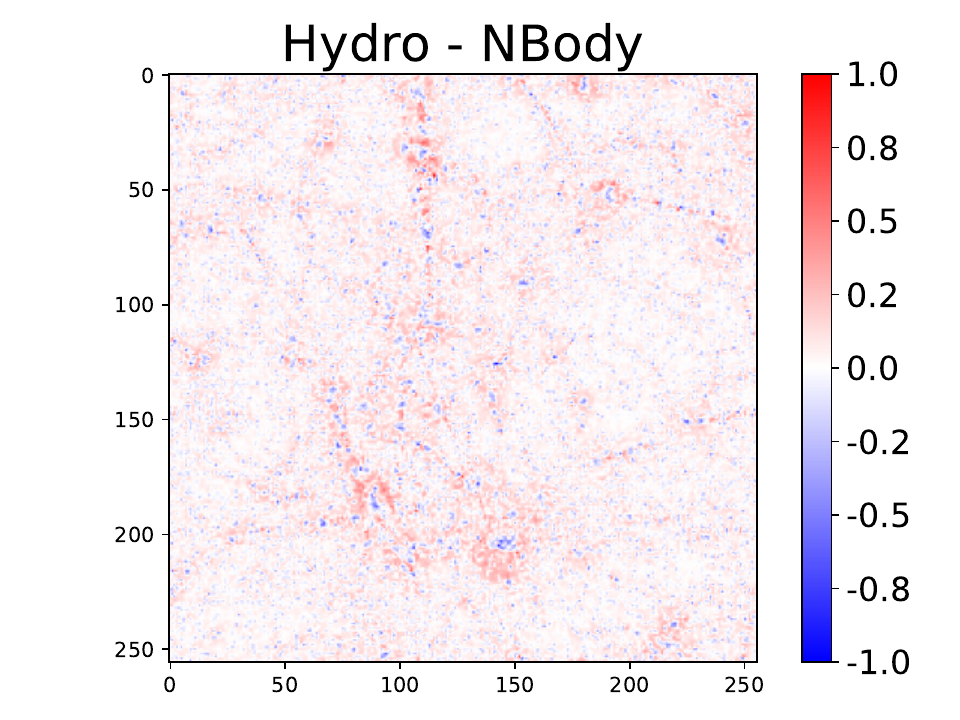}
\end{subfigure}

\begin{subfigure}{\columnwidth}
  \includegraphics[width=1.2\linewidth]{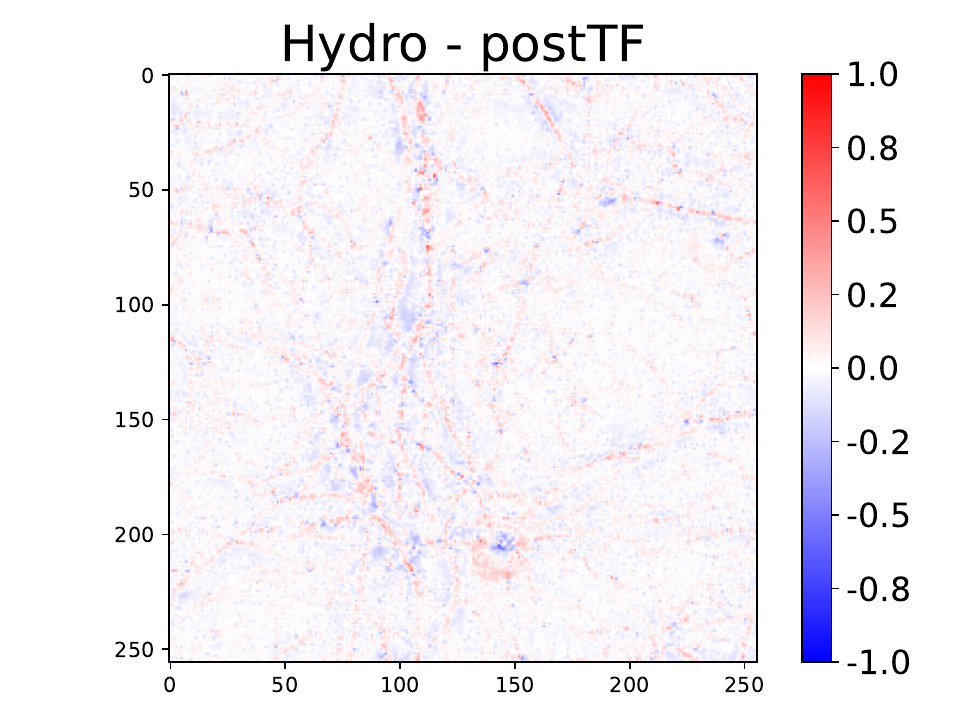}
\end{subfigure}

\begin{subfigure}{\columnwidth}
  \includegraphics[width=1.15\linewidth]{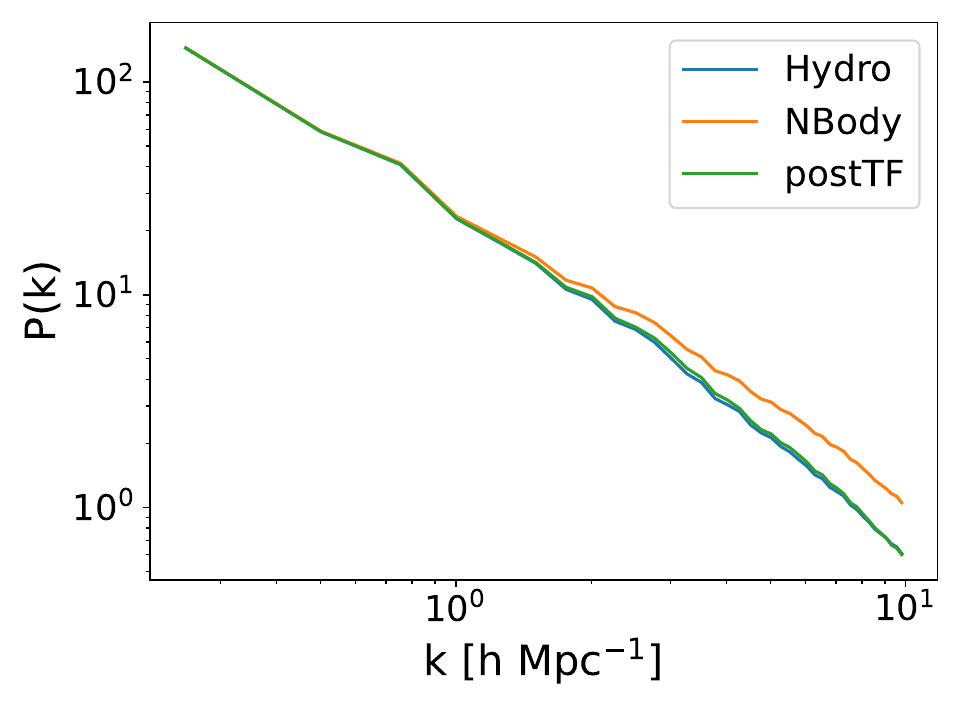}
\end{subfigure}

\columnbreak  

\begin{subfigure}{\columnwidth}
  \includegraphics[width=1.2\linewidth]{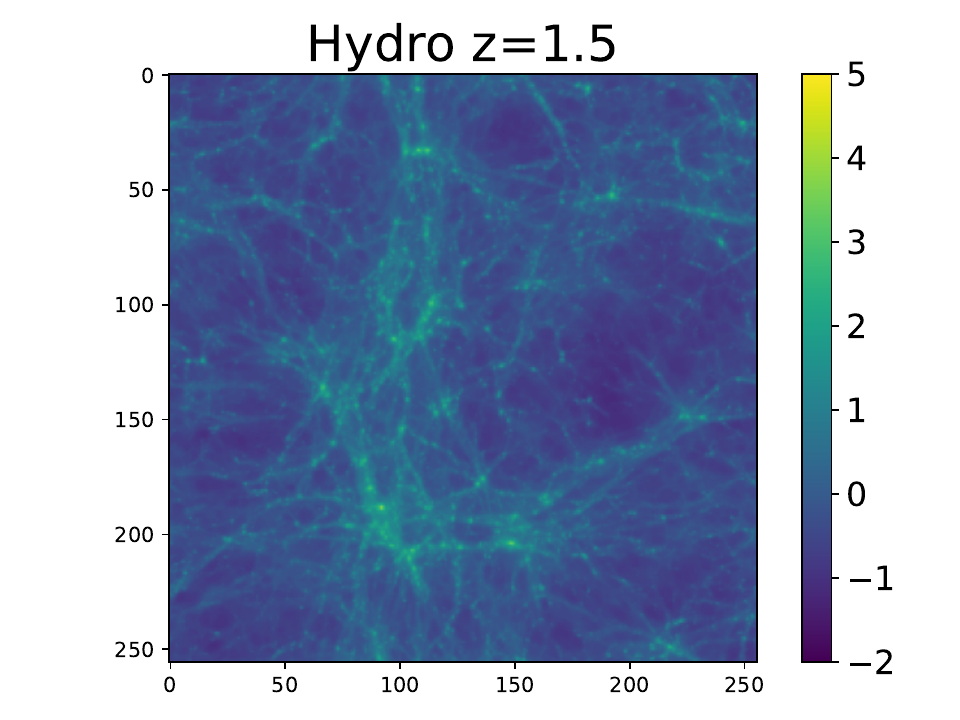}
\end{subfigure}

\begin{subfigure}{\columnwidth}
  \includegraphics[width=1.2\linewidth]{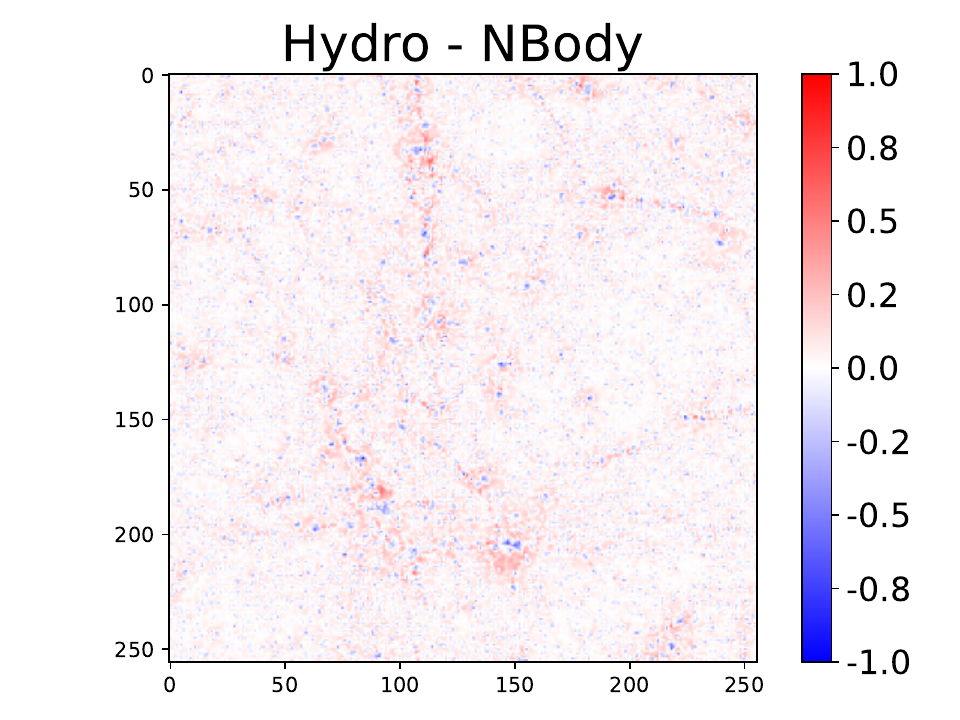}
\end{subfigure}

\begin{subfigure}{\columnwidth}
  \includegraphics[width=1.2\linewidth]{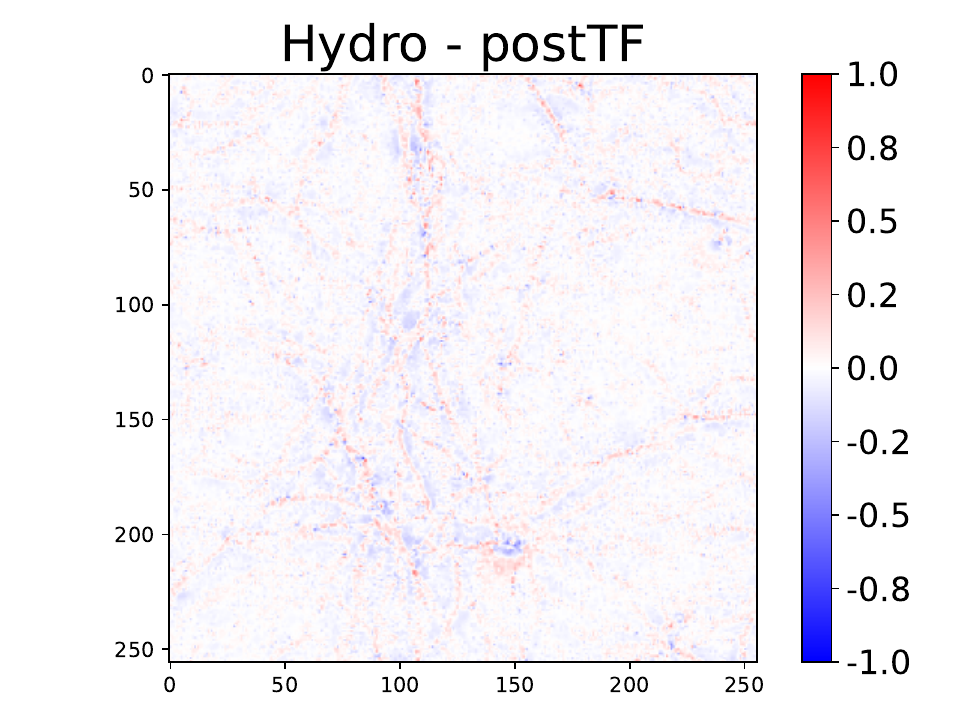}
\end{subfigure}

\begin{subfigure}{\columnwidth}
  \includegraphics[width=1.15\linewidth]{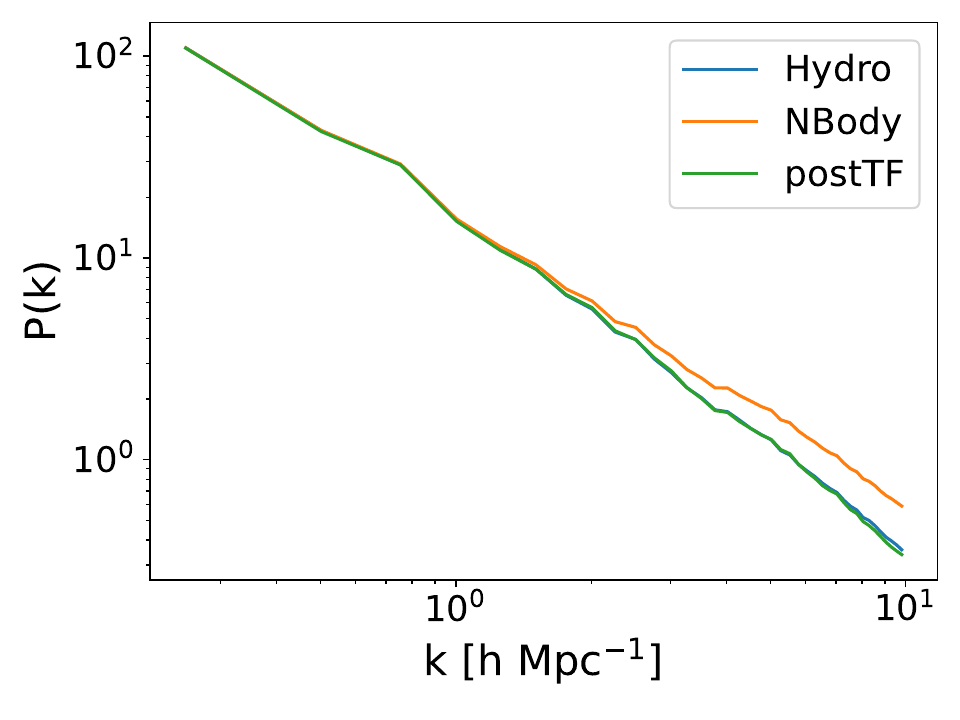}
\end{subfigure}

\end{multicols}
\caption{Field-level improvement comparison in the IllustisTNG simulation suite across different redshifts. The top row presents the original hydrodynamic field, while the bottom row shows the power spectra of the resultant N-body field after applying the transfer function, showcasing a substantial enhancement in power spectra alignment with the hydrodynamic fields. The achieved near-perfect calibration signifies improved field-level agreement, based on the implications from figures \ref{fig:CAMELSRatios} and \ref{fig:CAMELS_CC}. The second row displays the residual differences between hydrodynamic and initial N-body fields, whereas the third row demonstrates the residuals post-application of the transfer function (postTF). Notably, the post-transfer function residuals reveal a predominantly white background, indicating significantly improved agreement on both large scales and around clustered regions and halos on smaller scales. This illustrates the transfer function's efficacy in achieving notable field-level improvements, showcasing robustness across various redshifts.
}
\label{fig:IllustrisTNGFieldImages}
\end{figure*}

Figure \ref{fig:IllustrisTNGFieldImages} 
illustrates the baryonic correction of our method on IllustrisTNG when applied to N-body simulations across various redshifts. The first row displays a 2D projection of the whole 3D matter field with dimensions $25\times25\times 25~(h^{-1}{\rm Mpc})^3$ from a hydrodynamic simulation at four different redshifts. The second row shows the difference between the image from the hydrodynamic simulation and its N-body counterpart. The third row shows instead the difference between the hydrodynamic simulation and our field-level correction model. As expected, our field-level correction is more accurate than the N-body simulation, and the residual fluctuations (shown in red and blue) are due to small-scale modes where the cross-correlation coefficient deviates from 1.

\section{Conclusions}
\label{sec:Conclusions}

Field-level approaches have the potential to extract all the available information from cosmological surveys. Modeling and marginalizing over baryonic effects at the field-level becomes a key ingredient in these efforts. In this work we have developed a new method to model baryonic effects for the total matter density field, the relevant quantity for weak lensing analyses.

The key finding in this work is that by computing the cross-correlation between the total matter density field in hydrodynamic and N-body simulations from thousands of simulations of the CAMELS project (see Figure \ref{fig:CAMELS_CC}) we conclude that baryonic effects weakly affect the phases of Fourier modes of the total matter density field down to scales as small as $k\sim10~h{\rm Mpc}^{-1}$. 
This finding implies that baryonic effects will predominantly modify Fourier mode amplitudes. Thus, we can \textit{baryonify} the total matter field of an N-body simulation by rescaling its Fourier mode amplitudes. 

In this work we have built an emulator using Gaussian processes for the total to dark matter power spectrum ratio $S(k)$ that takes as input 2 cosmological parameters ($\Omega_{\rm m}$ and $\sigma_8$) and 4 astrophysical parameters ($A_{\rm SN1}$, $A_{\rm SN2}$, $A_{\rm AGN1}$, $A_{\rm AGN2}$). We have trained our emulator using 800 state-of-the-art hydrodynamic simulations from the Astrid suite of CAMELS. We then show that our emulator is able to reproduce the baryonic effects of thousands of hydrodynamic simulations that have different cosmologies, astrophysics, subgrid physics, resolutions, volumes, and redshifts within a few percent precision.

We have compared our emulator against others in the literature, such as BACCO, HMCode, and BCemu. We find that our emulator shares a similar level of accuracy with those, but it has a wider range of validity given that it has been trained on CAMELS, where variations in cosmology and astrophysics are very large. We also showed explicitly how using our method reduces the residuals when working at the field level by comparing the results of hydrodynamic simulations against \textit{baryonified} N-body simulations. A limitation of using CAMELS is that the box 
size is very small, and baryonic 
effects may not be fully captured 
due to the absence of larger 
halos in these simulation boxes.
This will need to be investigated 
in more detail using a suite 
of simulations varying box size. 

Our emulator enables robust, cost-effective field-level weak lensing modeling and facilitates precise power spectra analyses at the two-point level. The versatility and accuracy of our GP baryonification emulator underscore its potential as a powerful tool in cosmological simulations, offering opportunities for enhanced analyses and deeper insights into baryonic effects in large-scale structures. However, whether this emulator suffices at the field level depends on the specifics of the observational program. For example, for 
weak lensing, this will require making weak lensing maps using ray-tracing techniques. The
overall detectability of the 
effects that go beyond our 
field level emulator in the 
the weak lensing depends on  the density of 
background galaxies and the observed area of the sky. This analysis goes beyond the purpose of this paper, 
and will be presented elsewhere.

\section*{Acknowledgements}
We thank Raul Angulo and Aurel Schneider for their comments on the usage of
the BACCO and BCemu emulators. DS thanks James Sullivan for helpful discussions on Gaussian processes. The work of FVN is supported by the Simons Foundation. The CAMELS project is supported by the Simons Foundation and the NSF grant AST 2108078. 



\bibliographystyle{mnras}
\bibliography{references} 




\appendix


\bsp	
\label{lastpage}
\end{document}